# Effect of Alkyl-group Flexibility on the Melting Point of Imidazolium-based Ionic Liquids

Kalil Bernardino,[1*] Yong Zhang[2],

Mauro C. C. Ribeiro[1], Edward J. Maginn[2]

[1] *Laboratório de Espectroscopia Molecular*, *Departamento de Química Fundamental*, *Instituto de Química, Universidade de São Paulo, Av. Prof. Lineu Prestes 748, 05508-000, Brazil*

[2] *Department of Chemical and Biomolecular Engineering, University of Notre Dame, Notre Dame, Indiana, 46556, USA.*

*email: kalil.bernardino@gmail.com

# Effect of Alkyl-group Flexibility on the Melting Point of Imidazolium-based Ionic Liquids

The low melting point of room temperature ionic liquids is usually explained in terms of the presence of bulky, low symmetry and flexible ions, with the first two factors related to the lattice energy while an entropic effect is attributed to the latter. By means of molecular dynamics simulations, the melting points of 1-ethyl-3-methylimidazolium hexafluorophosphate and 1-decyl-3-methyl-imidazolium hexafluorophosphate were determined and the effect of the molecular flexibility over the melting point was explicitly computed by restraining the rotation of dihedral angles in both the solid and the liquid phase. The rotational flexibility over the bond between the ring and the alkyl chain affects the relative ordering of the anions around the cations and results in substantial effects over both the enthalpy and the entropy of melting. For the other dihedral angles of the alkyl group, the contributions are predominantly entropic and an alternating behavior was found. The flexibility of some dihedral angles has negligible effects on the melting point, while others can lead to differences in the melting point as large as 20 K. This alternating behavior is rationalized by the different probabilities of conformation defects in the crystal.

## I. INTRODUCTION

Ionic compounds are described in textbooks as having high melting points, existing as crystalline solids at room temperature. However, especially after 1990, several ionic compounds with melting points lower than 100ºC or even below room temperature were discovered and were called ionic liquids (ILs). In addition to displaying low melting point, ILs typically have very low vapor pressure and good conductivity at ambient conditions, characteristics that are desirable for applications like solvents, electrolytes in batteries, lubricants, and heat transfer fluids.[1-3]

Low melting points in ILs are usually achieved by the presence of bulky, low-symmetry and flexible ions. Bulky ions reduce the lattice energy due to the reduction in the electrostatic interaction between oppositely charged ions, thus decreasing the melting enthalpy. Low symmetry ions also result in less efficient packing in the crystalline phase, contributing to further reduction of the lattice energy. On the other hand, the presence of flexible ions is expected to reduce the melting point due to the increase of the melting entropy, since different conformers may exist in the liquid phase while only some conformations may be allowed in the solid.[1,4] However, even with those general guidelines, it is a challenging task to predict the melting point of a new IL or even whether it will be liquid at room temperature.[1,4] The presence and the strength of hydrogen bonds between the cations and anions, for instance, affect both the melting point and the viscosity of ILs and was investigated by both quantum and classical calculations.[5,6] In the present work, we focused on the question of how the flexibility of alkyl groups in imidazolium-based ILs affects the melting point. A brief review of some known effects of the flexibility over the crystallization, melting and other properties of ILs and other systems with amphiphilic ions is presented below.

An effect that is attributed to the conformational entropy of alkyl-chains in ionic liquids is the so-called “V-shape” seen in a plot of the melting point as a function of the number of carbon atoms, $n_C$, in the linear aliphatic chain of the cations. The melting point initially decreases with the increase of $n_C$ until a critical number after which there is a change of tendency and the melting point starts to increase with $n_C$.[7-9] The initial decrease is attributed to a dominant effect of the increase of conformational entropy in the liquid phase as the chain gets longer, while the increase for longer chains is explained as a predominance of the enthalpy resulting from van der Waals and hydrophobic interactions between chains.[8,9]

The possibility of different conformers results in the existence of different polymorphs in the solid phase that differ from each other mainly due to the conformation of some dihedral angle. The IL 1-ethyl-3-methyl-imidazolium dicyanamide, [C2C1IM][DCA], has two solid phases that differ from each other due to the in-plane or out-of-plane conformation of the cation ethyl group, as showed by Raman spectroscopy.[10] Computer simulations demonstrate that the conformation of the cation significantly changes the local spatial distribution of the anion.[8] The IL 1-butyl-3-methyl-imidazolium chloride, [C4C1IM][Cl], has a monoclinic crystalline form with a *gauche* dihedral between the first and the second carbon of the butyl group and an orthorhombic form with the same dihedral in the *trans* conformation.[11-13] It was suggested that the dynamics of conformational change in this liquid and in the corresponding bromide, [C4C1IM][Br], leads to supercooling and wide premelting.[14,15] Also, different water contents can change the preferred conformation of the cation in the crystal, with the conversion between different polymorphs taking several days in some cases.[16] For 1-butyl-3-methyl-imidazolium hexafluorophosphate, [C4C1IM][$PF_6$], three different polymorphs were characterized at different temperatures and pressures. Major differences include the conformation of the cations, the disorder of the anion and the packing density.[17] The protic ionic liquid propylammonium nitrate, $[C_3H_7NH_3][NO_3]$, presents the cation in a *trans* conformation at low temperatures but exhibits a transition to a crystal with *gauche* conformation upon heating.[18] Conformational freedom of ethyl chains was demonstrated to also be an important step during the pre-melting of triethylsulfonium bis(trifluoromethanesufonyl)imide, [S222][NTf2], upon heating.[19]

The flexibility of the anion also affects the melting point if it has multiple conformations. For example, bis(trifluoromethanesulfonyl)imide, [NTf2], is a bulky anion that has two conformers, the *trans* with point symmetry C2 and the *cis* with symmetry C1. [NTf2] is known to produce ionic liquids with low melting points.[1,4] It has been found that a conformational disordering of the anion takes place before melting, suggesting that the flexibility of the anion contributes to the low melting point.[20] Computer simulations showed different free energy surfaces for the two conformers of this anion around the [C2C1IM] cation.[21]

The introduction of ether functionalization in cation alkyl chains or perfluoralkyl chains enhances the flexibility of the chain, resulting in both lower viscosity[22,23] and lower melting points[23,24] than the pure alkyl counterparts.

If on one hand the flexibility affects the physical properties, the chemical environment can also change the conformation or the relative distribution of conformers. [NTf2] anion, for instance, forms crystals with different conformations depending on the cation.[25] Molecular simulations also showed that the relative populations of conformers of amphiphilic ions in solution changes due to the differences in the chemical environment. For instance, introduction of urea to an aqueous solution[26] or the transference of ions from a hydrophobic to a hydrophilic environment affects the population of different conformers.[27]

This brief review shows that the flexibility of both cations and anions is important for the comprehension of the melting point and possible crystalline structures of ILs. However, several questions remain. How much does the flexibility of a given dihedral angle affect the melting point? Or, in other words, if a given flexible dihedral angle is turned rigid, in the sense that it is trapped in a defined conformation, how much will the free energy, the entropy and the enthalpy difference between the liquid and the solid change? Does this effect depend on the position of the dihedral angle in a given alkyl chain? In the real world, one cannot control the flexibility of a dihedral angle without making chemical modifications like the ether functionalization discussed above or the introduction of double bonds. These modifications will introduce other effects alongside with the flexibility change like modifications in the dipole moment or bond lengths. On the other hand, in computer simulation it is possible to restrain a dihedral angle without changing the chemical composition in order to explore a pure flexibility effect. In the present work, we aim to answer these questions by the study of 1-ethyl-3-methyl-imidazolium hexafluorophosphate, [C2C1IM][PF6], and 1-decyl-3-methyl-imidazolium hexafluorophosphate, [C10C1IM][PF6], which will be referred to as [C2] and [C10], respectively, in the rest of this paper. First, the melting point of both ILs was determined using the pseudosupercritical path (PSCP) method.[28,29] Then the effect of flexibility was studied by restraining each dihedral angle one at a time and computing the free energy difference between the solid with and without restraint and between the liquid with and without restraint, which enables a direct calculation of the change of the melting point upon the removal of the flexibility of each dihedral angle.

## II. METHODS

### Software employed and simulation conditions

All the molecular dynamics simulations were performed using the LAMMPS package[30] with GAFF (General AMBER force field)[31] parameters and the same partial charges employed in previous works.[9,32] Figure 1 (top) gives the atom names that will be used in the rest of this paper. Note the difference in notation between atom C2 of the cation and system [C2], and the same for C10 and [C10]. The only parameters that were changed from previous works are the parameters for the CR2-N1-C1-C2 dihedral, which were found to give an incorrect description of the secondary minimum as well as an underestimated rotation barrier at 180º. It was reparametrized to fit the results of a potential energy scan done at the MP2/Def2-TVZPD level[33] performed with Orca 4.0.1.2.[34] The description of the reparameterization and the new parameters are given in Supporting Information. Due to this change, the results presented here differ slightly from the ones of previous works.[9,32] The unit cell for both [C2] and [C10] crystals[35] are composed of four ion pairs (Figure 1 bottom). To generate the simulation box for the solid phase, the unit cell of [C2] was replicated 5 times in the Y and Z directions and 4 times in the X direction while the unit cell of [C10] was replicated 4 times in each direction. This resulted in model systems with 400 and 256 ions pairs for [C2] and [C10], respectively. The initial structure for the liquid was prepared by heating the crystal rapidly from 250 to 600 K at a constant pressure of 1 atm, which ensures spontaneous melting for both compounds. The systems were cooled back to 300 K and the simulation boxes were deformed to a cube.

Structural analyses presented in this paper were performed with the TRAVIS[37] software and the graphical representations were rendered using VMD 1.9.3.[38]

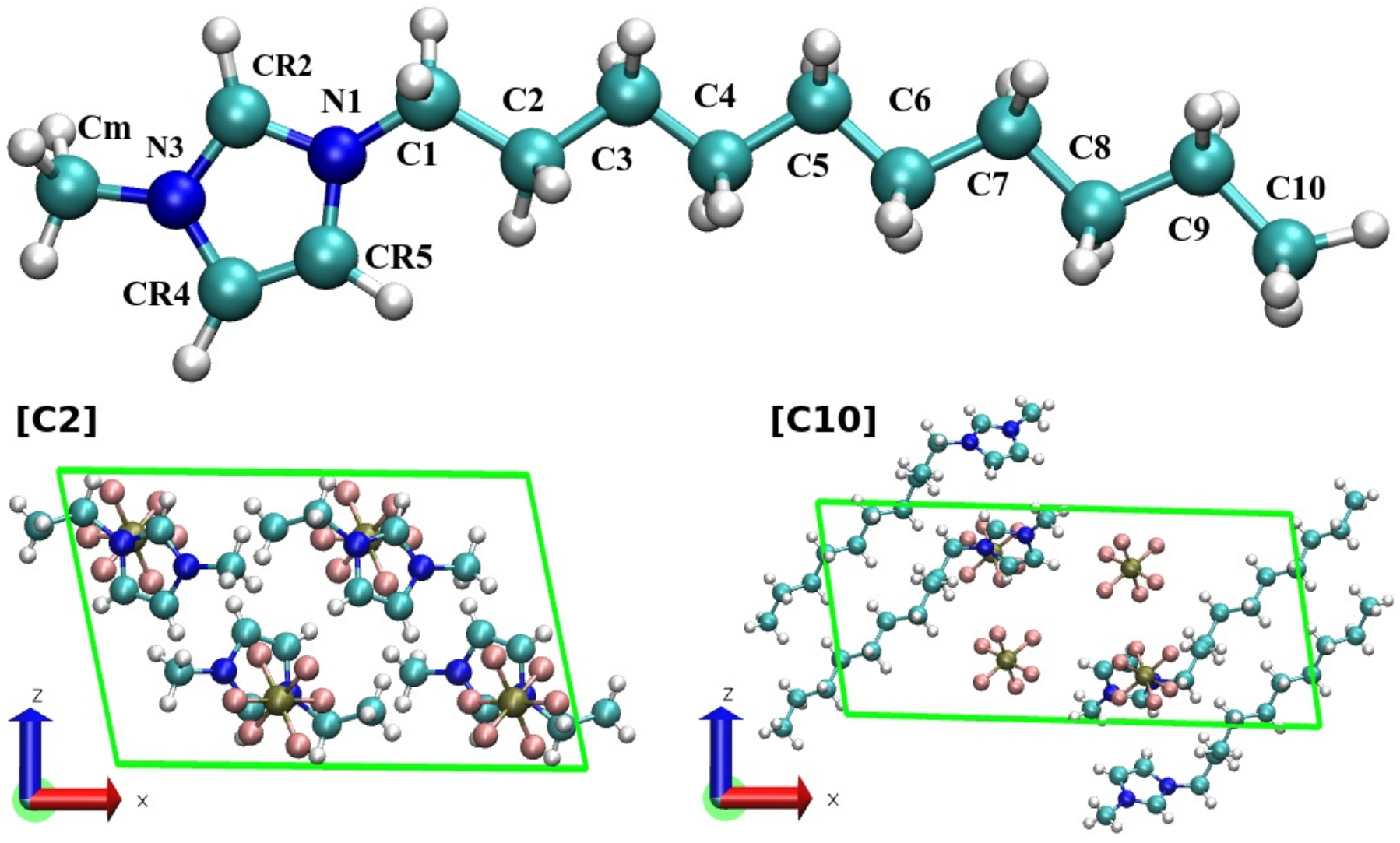

**Figure 1** Top: Structure of 1-decyl-3-methylimidazolium cation, [C10], with the atom names labeled. The same names were employed for the 1-ethyl-3-methyl-midazolium cation, [C2]. Bottom: Unit cells used to create the initial structure for the [C2] (left) and [C10] (right) crystals.

A cutoff radius of 12.0 Å was used for short-range Coulomb and Lennard-Joned-Jones interactions as well as for the tethering potential employed in the PSCP method, as described in next section. Ewald summation was used for long-range electrostatic interactions with a *k*-space resolution of 0.0001 and a tail correction was employed to account for long-range dispersion interactions. The simulations for the PSCP and the thermodynamic cycle to compute the effect of dihedral angle restraints were performed in the NVT ensemble while the simulations employed for the temperature scans were performed in the NPT ensemble with isotropic coupling for the liquid phases and anisotropic coupling for the solid phases. The Nosé-Hoover thermostat[36] with a damping parameter of 100.0 ps was employed in all the simulations, while the barostat with p=1.0 atm and damping parameter of 100.0 ps was used in the NPT simulations. A timestep of 1.0 fs was employed in all simulations.

**PSCP method**

The melting point would be largely overestimated if one simply tried to increase progressively the temperature of a bulk crystal in a molecular dynamics simulation. Direct phase coexistence methods are not reliable for computing the melting point of ionic liquids, due to the slow dynamics and complex reordering that occurs in going from the liquid to the crystal. Therefore, special methods are required to compute the melting point. In this work we employed the PSCP method[9,11,28,29,32,39] to compute melting points.

The PSCP method is comprised of four steps to obtain the free energy difference between the solid and liquid phases at a given temperature. These are schematically represented in Figure 2. In Step 1, thermodynamic integration is employed to weaken the non-bonded interactions of the crystal C (Equation 1).

$$U_1 = (1 - 0.9\lambda) U^{vdW} + (1 - 0.9\lambda)^2 U^{elect} + \lambda U^{teth} + U^{bonded}, 0 \leqslant \lambda \leqslant 1 \text{ (Eq. 1)}$$

At the same time, a tethering potential (Equation 2) is turned on to keep the ions vibrating around their lattice points, a state called "weak crystal", WC.

$$U^{teth} = \sum_i \sum_j a_{ij} \exp(-b_{ij} r_{ij}^2) \text{ (Eq. 2)}$$

Notice that the van der Waals interactions, $U^{vdW}$, are reduced to 10% of their original value while the Coulomb interactions, $U^{elect}$, are reduced to 1% of their original value when going from the C (λ=0) to the WC (λ=1) state. The potential for bonded interactions, $U^{bonded}$, was not changed along any of the thermodynamic integration steps. The tethering potential, $U^{teth}$, was applied over the C and N atoms of the cations and the P atoms of the anions. The parameter $a_{ij}$ was set to 16.0254 kJ/mol for the cation atoms and 14.0789 kJ/mol for the anion atoms, while $b_{ij}$ was set as 0.9 Å$^{-2}$ for every atom. It was shown in previous works that the choice of the tethering parameters do not affect significantly the computed free energy at the end of the PSCP cycle[29] as long as they are strong enough to hold the crystalline structure while reducing the strength of the non-bonded interactions.

The Helmholtz free energy difference between the C and the WC state was computed by numerical integration of the derivative of the system energy $U$ with the coupling parameter $\lambda$ (Equation 3) using the trapezoidal method.

$$\Delta A = \int_0^1 \left\langle \frac{\partial U}{\partial \lambda} \right\rangle d\lambda \quad \text{(Eq. 3)}$$

In Step 2, the tethering potential is turned off gradually using the following expression

$$U_2 = 0.1\, U^{vdW} + 0.01\, U^{elect} + (1-\lambda) U^{teth} + U^{bonded}, 0 \leqslant \lambda \leqslant 1 \quad \text{(Eq. 4)}$$

It is during this step that the melting of the crystal happens. The amorphous state obtained at the end of this step is called *"weak dense fluid"*, WDF. The free energy change for step 2 is computed again by the integral of $\partial U/\partial \lambda$ (Equation 3).

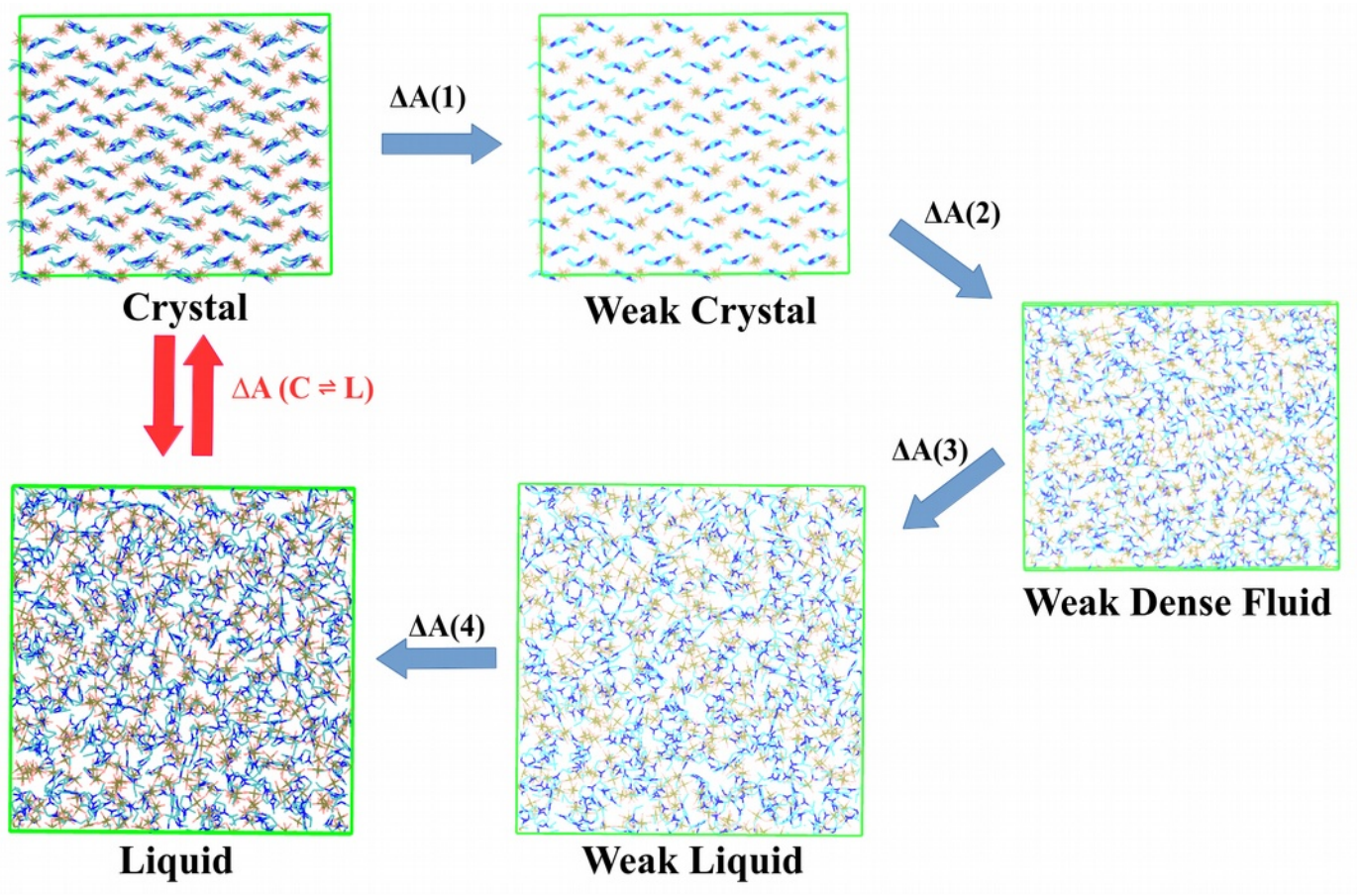

**Figure 2** Illustration of the PSCP method to compute the melting point.

In step 3, the volume of the system is changed from the crystal density to the liquid density while holding the intermolecular interactions at the same level as at the end of step 2. The resulting state is called the *"weak liquid"*, WL. The free energy difference between the WDF and WL states is computed via

$$\Delta A(3) = -\int_{V_C}^{V_L} p dV \quad \text{(Eq. 5)}$$

where $V_C$ and $V_L$ are the average volumes of the crystal and liquid obtained in NPT simulations, respectively, and $p$ is the measured pressure of the system.

In Step 4, the non-bonded interactions are restored to their full strength

$$U_4=(0.1+0.9\lambda)U^{\mathrm{vdW}}+(0.1+0.9\lambda)^2U^{\mathrm{elect}}+U^{\mathrm{bonded}}, 0\leqslant\lambda\leqslant 1 \quad \text{(Eq. 6)}$$

and the liquid, L, is obtained. The Helmholtz free energy variation for this step is obtained again by the integral in Equation 3 using $U_4$ instead of $U_1$.

The melting point corresponds to the temperature at which the difference in the Gibbs free energy between the solid and liquid phases is zero. To compute the Gibbs free energy at the temperature that the PSCP cycle was performed, the $\Delta(pV)$ term between the two phases is added to the sum of Helmholtz free energies computed in the PSCP cycle (Equation 7).[11]

$$\Delta G_{\mathrm{ref}}=\Delta A(1)+\Delta A(2)+\Delta A(3)+\Delta A(4)+p(V_L-V_C) \quad \text{(Eq. 7)}$$

The Gibbs free energy difference as a function of the temperature was computed from the $\Delta G_{\mathrm{ref}}$ obtained at the reference temperature $T_{\mathrm{ref}}$ employed in PSCP and the Gibbs-Helmholtz equation

$$\frac{\Delta G(T)}{T}-\frac{\Delta G_{\mathrm{ref}}}{T_{\mathrm{ref}}}=-\int_{T_{\mathrm{ref}}}^{T}\frac{\Delta H(T)}{T^2}\,dT \quad \text{(Eq 8)}$$

where $\Delta H(T)$ is the enthalpy difference between the liquid and solid, which can be obtained directly from the simulation of both phases at different temperatures.

The enthalpy of both phases and for both the [C2] and [C10] increases linearly with temperature until spontaneous melting is observed for the solid at high temperatures (Figure S2 in Supporting Information). Discarding the data points above the spontaneous melting, the following linear relation was fit for the enthalpy change

$$\Delta H(T)=H_L(T)-H_C(T)=\Delta H_0+\Delta C_p T \quad \text{(Eq 9)}$$

where the slope is the difference of heat capacity between the two phases, $\Delta C_p$. This linear relation for $\Delta H(T)$ was employed to solve the Gibbs-Helmholtz equation resulting in the closed form equation for the free energy difference between the solid and liquid phases as a function of temperature

$$\frac{\Delta G(T)}{T}=\Delta H_0\left(\frac{1}{T}-\frac{1}{T_{\mathrm{ref}}}\right)-\Delta C_p\ln\left(\frac{T}{T_{\mathrm{ref}}}\right)+\frac{\Delta G_{\mathrm{ref}}}{T_{\mathrm{ref}}} \quad \text{(Eq 10)}$$

Simulations of 4 ns were carried out for each λ in steps 1 and 4 and for each volume in step 3, while 6 ns simulations were performed for each λ in step 2 due to the slower convergence for the λ values in which the transition from the structured to the amorphous state was observed. In each case, the first half of the simulations were discarded in order to guarantee equilibration.

**Dihedral angle restraints**

To access the effect of molecular flexibility on melting point, rigidity was added by means of an extra restraining potential applied over the dihedral angles, defined by

$$U_{\mathrm{restr}}(\phi)=\lambda K\left[1+\cos(\phi-d)\right] \quad \text{(Eq 11)}$$

where $0 \leq \lambda \leq 1$ is a coupling parameter, $\phi$ is the dihedral angle and $d$ equals 180º minus the target angle chosen as the most probable angle in the distribution of the dihedral angle in the crystal without restraints. This restrained potential is used in addition to the normal force field parameters, but with a stronger force constant, $K$ = 83.7 kJ/mol. This $K$ is strong enough to enable only small oscillations around the minimum of the potential defined in Equation 11 at the full strength ($\lambda$=1).

In principle, the same PSCP cycle employed for the unrestrained molecules could be employed for the restrained molecules. However, we used a different thermodynamic cycle to obtain the free energy difference between the new liquid and the new crystal with the restrained dihedral angles, which will be called “Restrained Liquid”, RL, and “Restrained Crystal”, RC (Figure 3). In this cycle, in step a, the restrained potential is applied progressively over the crystal C by changing the coupling parameter $\lambda$ from 0 to 1 in Equation 11, and the free energy variation of this step, $\Delta$A(a), is computed by numerical integration of Equation 3. This step is done in the NVT ensemble, and the final state (called RC*) has the same volume as the unrestrained crystal. A volume change is then performed in step b to go from the volume of C to that of RC. The corresponding free energy, $\Delta A$(b), is computed by the integral of $-p\mathrm{d}V$.

The same procedure described for the solid is applied to convert L to RL in steps c and d. The free energy difference between C and L is obtained from the PSCP method, which completes the cycle. The convergence of the simulations along this new cycle is faster than the PSCP cycle with smaller statistical error, as will be shown in the Results and Discussion section. Only 2 ns simulations were used for each $\lambda$ in steps a and c or for each volume in steps b and d. Another advantage of introducing this new cycle lies in the interpretation of the results. With this new protocol, one can access directly the effect of the restraints over each phase separately, which would not be possible if the PSCP cycle were employed for the restrained phases.

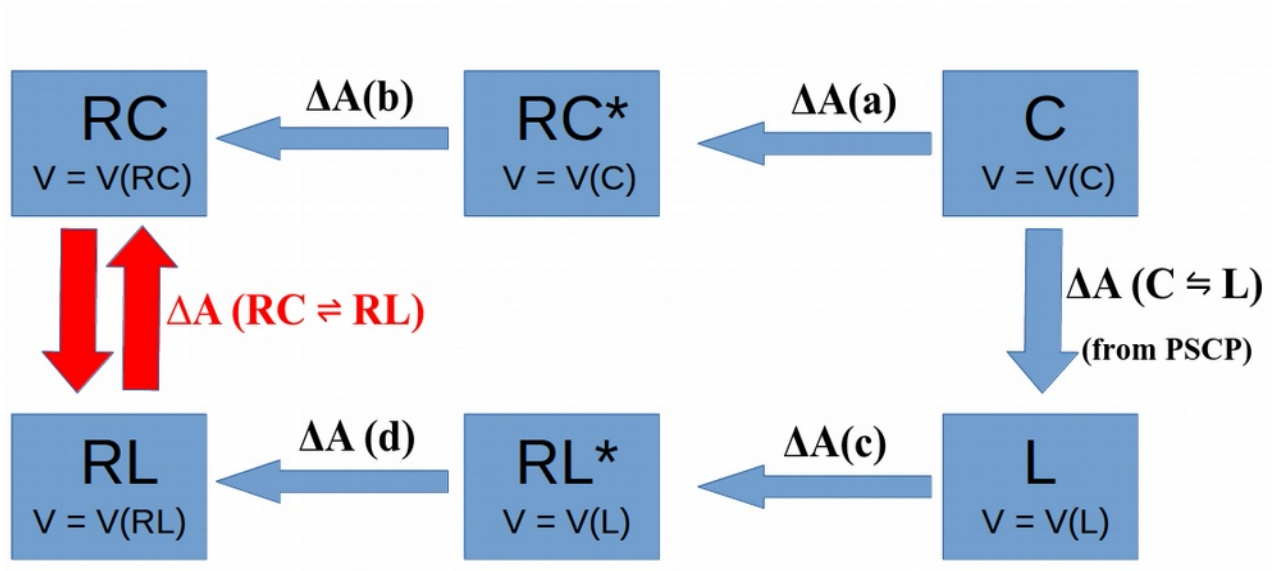

**Figure 3** Thermodynamic cycle used to compute the melting free energy in the system with a restrained dihedral. C and L denote the crystal and the liquid without restraints, RC* and RL* the crystal and liquid with the restraints at the same volume of the unrestrained systems, and RC and RL represents the phases with restraints at their respective equilibrium volumes.

As in the PSCP method, the $\Delta(pV)$ term needs to be added to convert the Helmholtz free energies into the Gibbs free energy

$$\Delta G_{ref,R} = \Delta A_{PSCP} + \Delta A(c) + \Delta A(d) - \Delta A(a) - \Delta A(b) + P(V_{RL} - V_{RC}) \quad \text{(Eq 12)}$$

The temperature dependence of enthalpy was determined by performing NPT simulations of the RC and RL as in the case of unrestrained systems. Linear relations were found and Equation 10 was used to determine the free energy difference between RL and RC as a function of temperature.

The value of 83.7 kJ/mol for the constant $K$ in the restraint potential is arbitrary. However, the curves $\partial U/\partial \lambda$ for the solid and the liquid converge to the same values before $\lambda = 0.3$ for every dihedral angle studied (Figures S11 to S20 in Supporting Information), which suggests that any $K$ value larger than 25.0 kJ/mol would result in the same free energy difference between RC and RL phases and the same melting point. A value smaller than 25.0 kJ/mol, however, could permit the existence of conformation defects that we want to eliminate by the use of the restraint potential.

**Statistical error calculation**

The statistical error over the free energy values computed during each step of the thermodynamic cycles was estimated by dividing the equilibrium part of the trajectories into four equal size blocks. Free energy change was calculated using one block of data for each simulation, resulting in four different free energy values for each step. The standard deviation between these blocks was taken as the statistical error for the given step. The statistical error in the free energy difference between C and L and between RC and RL were computed in a similar way using only one block for each step along the cycle (Equations 7 and 12), thus obtaining four different results for the free energy which enables the calculation of the standard deviation between them. Each of the NPT simulations used to determine the enthalpy variation with temperature was also divided into four blocks. The linear regression defined in Equation 9 was applied to each set of blocks, resulting in four values of $\Delta C_p$ and $\Delta H_0$ for each system. The standard deviation was then calculated. Finally, using four sets of $\Delta G_{ref}$, $\Delta C_p$ and $\Delta H_0$, four melting points were obtained and the standard deviation was calculated in a similar manner.

## III. RESULTS AND DISCUSSIONS

**Flexible systems**

In this section, we will discuss the results for the systems without the dihedral angle restraints, which not only serves as a baseline to discuss the flexibility effect but the free energy obtained from the PSCP method for the flexible molecule (Figure 2 and Equation 7) is also a part of the thermodynamic cycle used for the restrained systems (Figure 3 and Equation 12).

Both [C2] and [C10] crystals do not melt spontaneously when performing NPT simulations at the reference temperatures of 380 and 425 K, respectively, but significant conformational disorder of the ethyl and decyl groups was observed (see discussion in the next sections) as well as rotational disorder of the $[PF_6]^-$ anions. Both effects are expected, since similar disorder was observed experimentally even below the melting point for the homologous IL [C4C1IM][PF6][17] and other IL.[20] However, a spontaneous structural change took place for [C10] during the initial relaxation. The N1-C1-C2-C3 and C2-C3-C4-C5 dihedrals were reported to exist as a predominant *gauche* conformation in the crystal[33] (Figure 1, bottom). However, in the simulation, the system quickly evolved to a different structure in which the *trans* conformation is preferred. To guarantee that this is not due to improper initial equilibration, we tried to relax the crystal structure at 300 K, which is lower than the experimental melting point. We also tried to apply a restraint to hold these dihedral angles at the initial conformation during the first equilibration simulation and then removed the restraints and performed a second equilibration. In both tests the crystal evolved to a structure in which every dihedral angle presented the *trans* as the most favorable conformation. This deviation from experimental crystal structure may be because the force field overstabilizes the *trans* conformers, an effect observed in a previous work for a different IL with a different force field.[9] It is also possible for [C10] to take a crystal structure at higher temperatures that is different than that determined at 173 K.[33] Despite this divergence, we proceed with the calculations for this model system and focus on the general features of the dihedral angle flexibility of alkyl groups over the melting of IL.

Figure 4 shows the simulation results for each step along the PSCP cycle for both [C2] (red curves) and [C10] (black curves). The $\partial U/\partial \lambda$ changes sign during step 1 due to the balance between the non-bonded interactions, that become progressively weaker, and the tethering potential, that becomes stronger as λ increases (Equation 1) and eventually becomes dominant over the non-bonded interactions. For [C10], the sign change happens at lower λ and the final values of $\partial U/\partial \lambda$ are more negative than those for [C2]. As a consequence, the resulting contribution of step 1 per ion pair for the free energy is smaller for [C10] than for [C2] (Table 1).

A crystalline to amorphous transition occurs during step 2, as can be seen by the drop in $\partial U/\partial \lambda$ between λ=0.7 and λ=0.8 in Figure 4. This step is the one that involves the largest statistical errors and presents slower convergence especially in the region of the transition. A large number of simulations with different $\lambda$ were carried out at the transition region to ensure an accurate numerical integration. Due to the large number of ions subjected to the tethering potential, the contribution of step 2 is larger for [C10] than for [C2] and is positive in both cases (Table 1), since attractive interactions are being removed from the system when turning off the tethering potential.

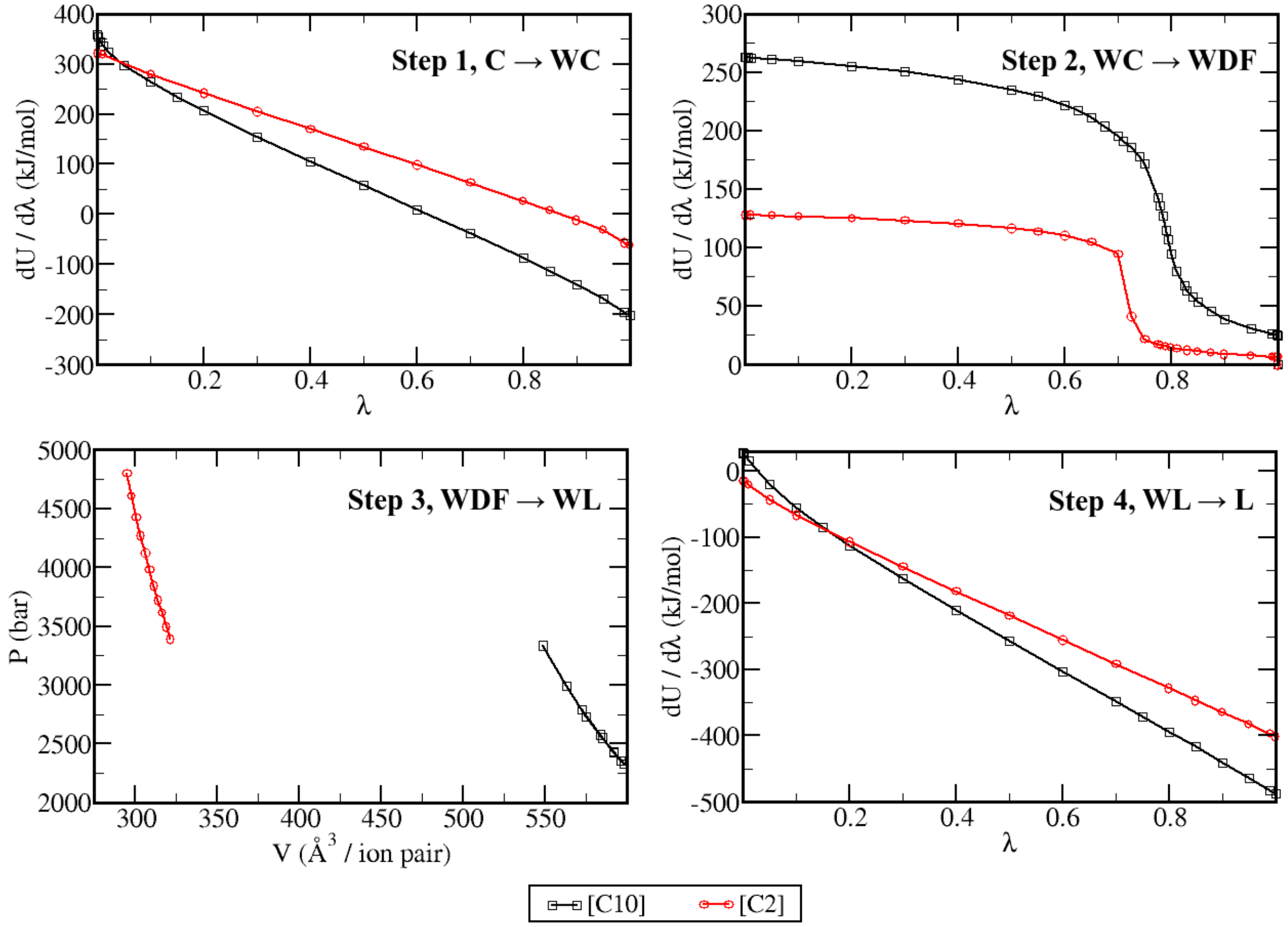

**Figure 4** Derivative of the internal energy *U* per ion pair with the coupling parameter *λ* for steps 1, 2, and 4 along the PSCP cycle and pressure as a function of volume per ion pair for step 3.

In step 3, the box dimensions are changed from the crystal phase to liquid state. Notice that the average pressure is huge at this step since the intermolecular interactions are weakened and the systems tend to expand. The change in the molar volume is larger for [C10], but the pressure is larger for [C2] and the overall contribution for the free energy of this step is more negative for [C2]. In step 4, the Coulombic and van der Waals interactions are restored leading to the liquid state. This results in a large negative free energy change (Table 1) that counteracts most of the changes in steps 1 and 2. Notice that the contribution of step 3 is one or two orders of magnitude smaller than the steps that involve changes in interaction potentials, but is not negligible. The overall free energy changes along the PSCP cycle are of the same order of magnitude of step 3.

**Table 1** Results from PSCP method for [C2] at 380 K and [C10] at 425 K. Values are in kJ/mol per ion pair. The subscript indicates the statistical error in the last digit. For example, $60.30_4$ means 60.30 ± 0.04.

| System | [C2] | [C10] |
|---|---|---|
| $\Delta A$ step 1 (C→WC) | $134.05_3$ | $60.30_4$ |
| $\Delta A$ step 2 (WC→WDF) | $88.23_5$ | $194.86_8$ |
| $\Delta A$ step 3 (WDF→WL) | $-6.552_4$ | $-4.36_1$ |
| $\Delta A$ step 4 (WL→L) | $-216.99_4$ | $-251.89_3$ |
| $\Delta(pV)$ (C→L) | $0.0020_1$ | $0.0014_1$ |
| $\Delta G_{ref}$ (C→L) | $-1.26_6$ | $-1.1_2$ |

Finally, the $\Delta(pV)$ between the crystal and liquid states was calculated, which makes negligible contribution to the total free energy. The free energy change is negative at the reference temperature for both compounds. This shows that melting should be spontaneous at 380 K for [C2] and at 425 K for [C10]. In the NPT simulations, melting of crystal was observed only at higher temperatures (Figure S2), showing the need to employ special methods like the PSCP to determine the melting point. Using equations 8 and 10 and results from the PSCP calculations, the melting temperatures were found to be 371 K for [C2] and 409 K for [C10], respectively (Figure 5).

The predicted melting points for both compounds are higher than the experimental values (Table 2). Since the focus in this work is to explore the effects of dihedral angle flexibility, we will use these values as a baseline for further discussion and explore the possible reasons for the overestimation in a separate publication.

**Table 2** Melting points and changes in enthalpy, entropy, heat capacity, and energy components at the melting temperature without and with the restraint over the CR2-N1-C1-C2 dihedral angle. Values are in K for $T_m$, J K$^{-1}$ mol$^{-1}$ for $\Delta S_m$, kJ K$^{-1}$ mol$^{-1}$ for $\Delta C_{p,m}$ and kJ/mol for enthalpy and energy components. Experimental data for $T_m$, $\Delta H_m$ and $\Delta S_m$ are given in parenthesis.

| System | [C2] flexible | [C10] flexible | [C2] restrained | [C10] restrained |
|---|---|---|---|---|
| $T_m$ | $355_1$ (334,[8] 336[40], 339[42]) | $409_2$ (307,[8] 308[41]) | $430_2$ | $405_2$ |
| $\Delta H_m$ | $19.3_1$ (17.7,[8] 17.9[42]) | $21.1_3$ (19.4[8]) | $24.9_1$ | $28.3_1$ |
| $\Delta S_m$ | $51.6_3$ (53.2,[8] 52.8[42]) | $51.4_8$ (63.3[8]) | $57.9_2$ | $69.8_8$ |
| $\Delta C_{p,m}$ | $0.007_2$ | $-0.034_4$ | $0.042_2$ | $-0.065_5$ |
| electrostatic | $10.24_4$ | $10.61_4$ | $12.80_2$ | $12.12_2$ |
| van der Waals | $7.87_2$ | $6.97_5$ | $10.54_4$ | $12.3_1$ |
| bonds | $0.106_4$ | $0.40_1$ | $0.025_1$ | $0.487_1$ |
| angles | $0.38_2$ | $1.99_4$ | $0.257_2$ | $1.94_1$ |
| dihedrals | $0.53_1$ | $1.02_2$ | $0.96_1$ | $1.32_2$ |

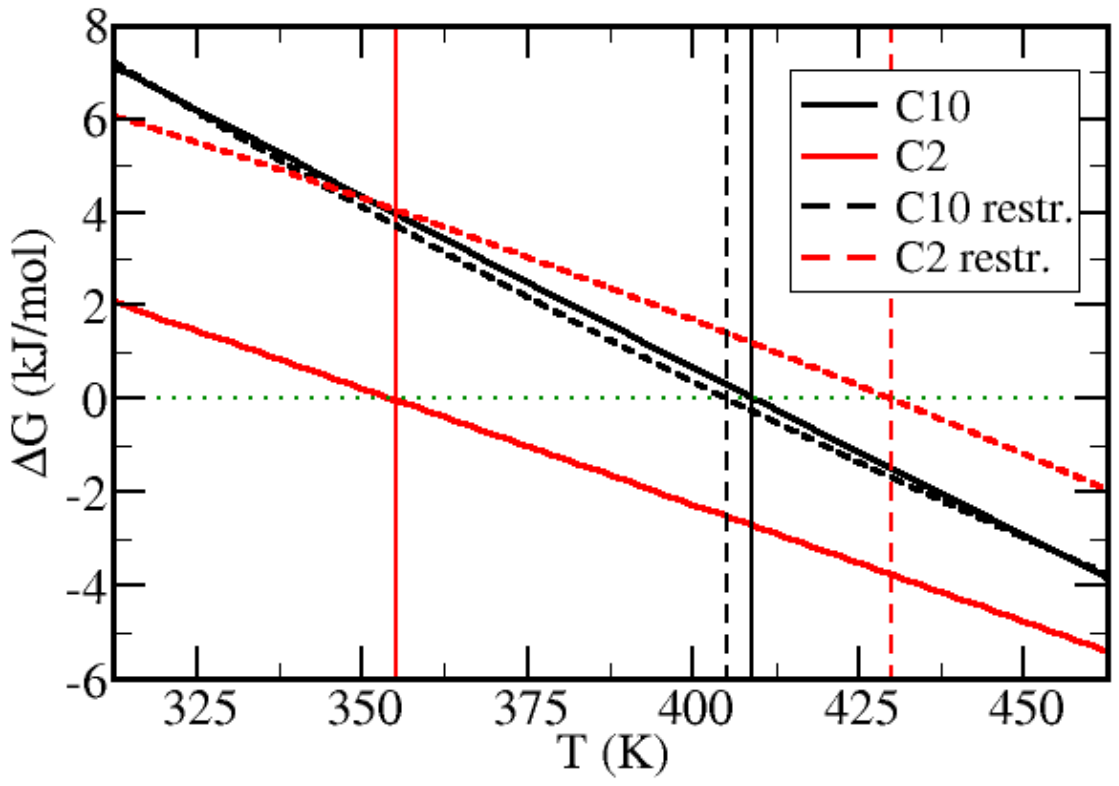

**Figure 5** Calculated free energy difference between the liquid and crystal phases. The solid lines stand for the systems without dihedral angle restraints and the dashed lines for the ones with restraint over the CR2-N1-C1-C2 dihedral angle. The vertical lines indicate the computed melting points.

**Effect of CR2-N1-C1-C2 dihedral**

The effect of dihedral angle flexibility on melting point was studied by the introduction of an extra restraint potential on each dihedral angle at a time (Equation 11). The effect on the CR2-N1-C1-C2 dihedral angle is discussed first due to its different distribution compared to other dihedral angles in the alkyl chain. Without restraints, the maximum distributions of the CR2-N1-C1-C2 dihedral angle in [C2] crystal is located close to ± 90° at high temperature and shifts to larger angles at low temperatures (Figure 6, solid lines stands for the crystal and dashed lines for the liquid). Another maximum occurs at 0° in the liquid phase for both compounds and the solid phase of [C10], as expected due to the dihedral angle potential energy surface (Figure S1). For [C2], the differences between the liquid and solid phases are remarkable. In the solid, there is no population at 0° and the distributions at ± 90° gets broader as temperature increases; in the liquid phase, there are three populations with significant superposition, which are independent of temperature. For [C10], on the other hand, the distributions for the solid and liquid phases are very similar. Except for the solid at low temperatures, both phases displayed a small maximum at 0°. The difference between the two compounds is likely related to the different role of van der Waals interactions among the aliphatic tails. The decyl groups in [C10] may maintain a similar cluster structure in both phases while the same effect is not present in [C2]. It is known that the "V-shape" feature in the melting point of imidazolium based ILs is due to the balance between enthalpy and entropy.[7] Relatively, the entropic effect dominates the melting of ILs with small alkyl chains and the enthalpic effect overcomes the entropic effect for long alkyl group ILs. Different roles of the alkyl group over the structure may also be expected for these two compounds.

This dihedral angle has two equally probable populations at *ca.* ± 90°. Dihedral angle restraint (Equation 11) was applied to hold half of cations at the positive angle positions and the other half at the negative angle positions in both liquid and crystal phase. The associated free energy change was calculated along the cycle defined in Figure 3 and the results are summarized in Table 3.

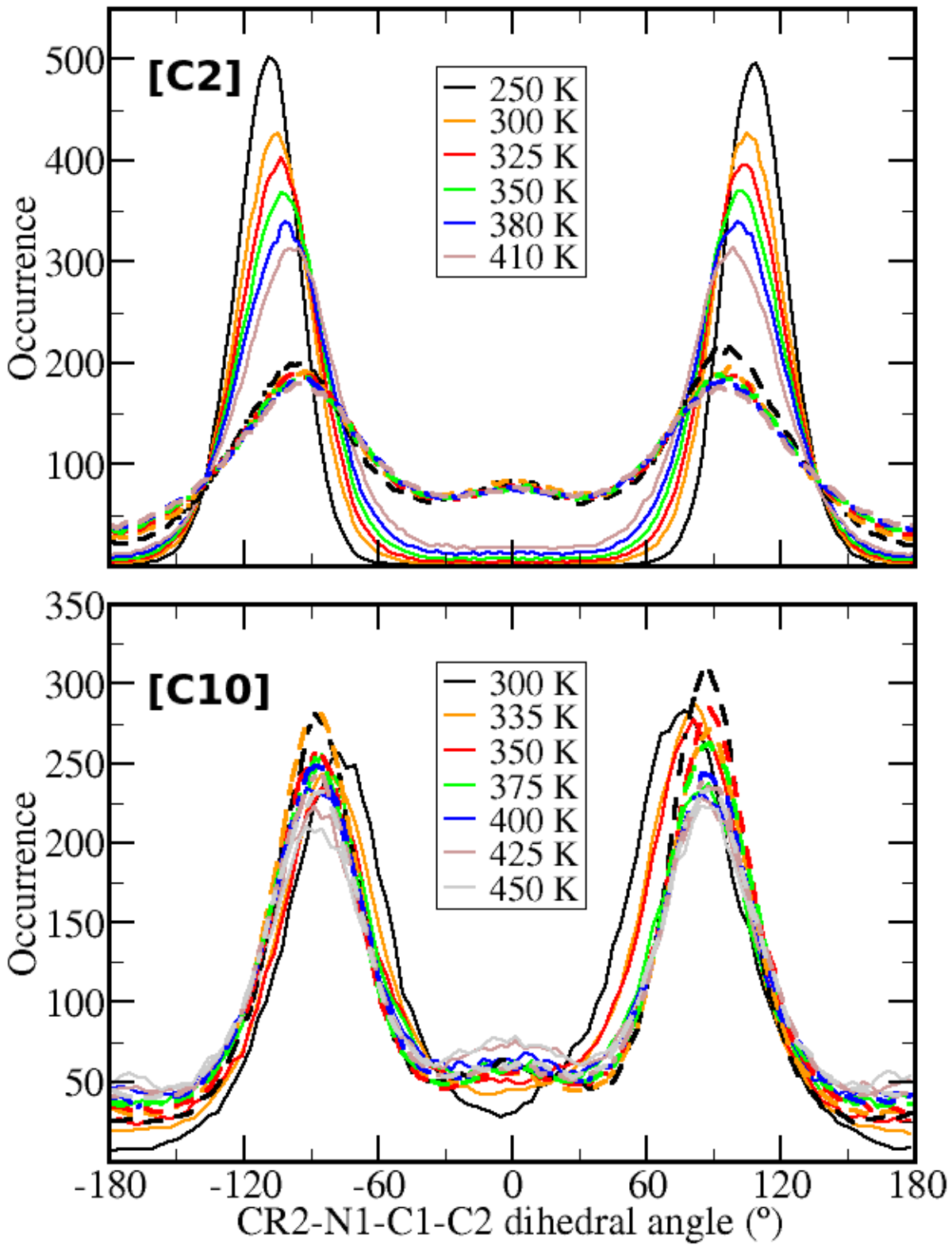

**Figure 6** Distribution of the dihedral angle CR2-N1-C1-C2 in [C2] (top) and [C10] (bottom) at different temperatures without additional dihedral restraints. Solid lines are the distribution in the crystal and dashed lines for liquid.

Introduction of the dihedral angle restraint produces similar free energy changes for both compounds in the liquid phase (step c in Table 3). Compared to the liquid phase, the effect of restraining this dihedral angle is much smaller in the solid phase for [C2] and similar for [C10]. Detailed data are given in Figure S11 for [C2] and Figure S12 for [C10] in Supporting Information. The free energy changes due to volume change are negligible in both compounds. As a result, the calculated $\Delta G_{ref}$ for [C10] with the dihedral angle restraint is very similar to that without the restraints (Tables 1 and 3) and the melting point was found to be 4 K lower, close to the statistical precision limit. For [C2], the restraint causes the $\Delta G_{ref}$ to increases by more than 4.0 kJ/mol and the melting point to increase by 75 K (Table 2 and Figure 5). Note that the statistical errors related to the steps along the cycle shown in Figure 3 (Table 3) are smaller than those along the PSCP cycle (Table 1). This demonstrates the advantage of using the cycle in Figure 3 instead of PSCP for the restrained systems.

The different behavior of [C2] and [C10] upon dihedral angle restraint can be understood by computing the melting entropy and enthalpy. The melting enthalpy can be further decomposed into bonded and non-bonded energy contributions (Table 2). Upon imposing a dihedral angle restraint, $\Delta H_m$ increases by 5.6 and 7.3 kJ/mol for [C2] and [C10], respectively, which means that the restraint stabilizes the solid relative to the liquid from an enthalpic point of view. At the same time, $\Delta S_m$ increases by 6.3 and 18.4 J $K^{-1}$ $mol^{-1}$, respectively, counteracting the enthalpy change and destabilizing the crystal. For [C2], the enthalpy effect is dominant and the melting point increases. For [C10], due to the larger entropy change, the effects of enthalpy and entropy cancel out each other, such that the restraint has a small effect on the melting point.

**Table 3** Calculated Helmholtz free energy change along the thermodynamic cycle of Figure 3 for the restraint over the CR2-N1-C1-C2 dihedral angle. Free energy values are in kJ/mol, melting points are in K. The subscripts indicate the computed uncertainty by block average in the last digit.

| System | [C2] | [C10] |
|---|---|---|
| $\Delta A$ step a (C→RC*) | $2.59_1$ | $6.51_3$ |
| $\Delta A$ step b (RC*→RC) | $0.005_1$ | $-0.05_1$ |
| $\Delta A$ step c (L→RL*) | $6.65_1$ | $6.31_3$ |
| $\Delta A$ step d (RL*→RL) | $-0.009_2$ | $-0.084_4$ |
| $\Delta(pV)$ (RC→RL) | $0.0013_2$ | $0.0002_1$ |
| $\Delta G_{ref}$ (RC→RL) | $2.91_9$ | $-1.4_1$ |
| $T_m$ | $430_2$ | $405_2$ |

Further analysis reveals that the change in the melting enthalpy arises mostly from the solid phase, and the change in the electrostatic component implies that the organization of the ionic parts is affected by the dihedral angle restraint. This is confirmed by the analyses of radial pair distribution function between atoms of the cation and anion (Figure 7). In the liquid phase (dashed lines), there is only a small change of the structure limited to the first coordination shell. For the solid phase, the structural change due to the dihedral angle restraint is more intense and propagates to the second and further coordination shells (solid lines).

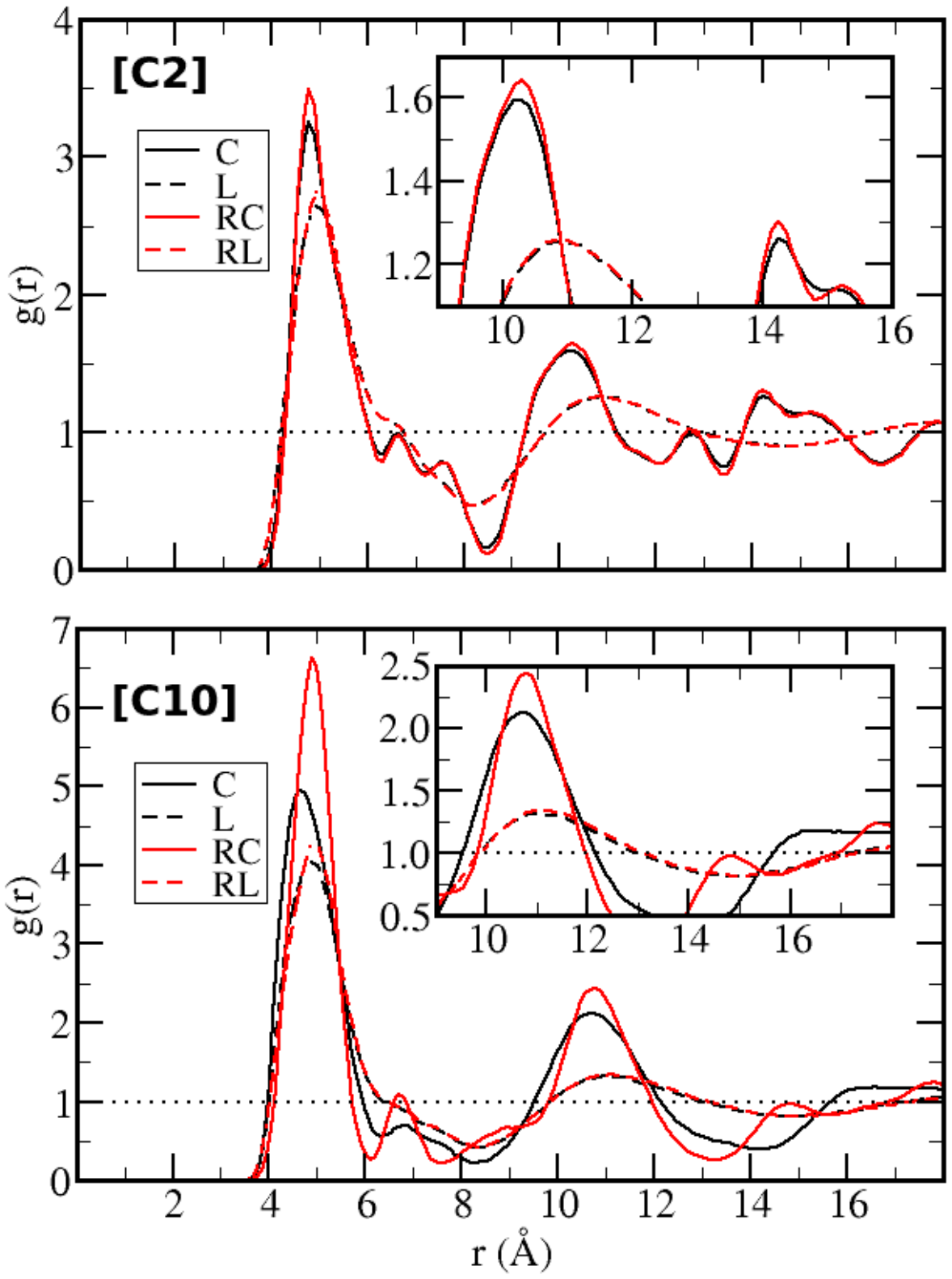

**Figure 7** Radial distribution function, $g(r)$, between the N1 atom of the cation and P atom of the anion for [C2] at 380 K (top) and for [C10] at 425 K (bottom). The insets are zoom-ins of the second and third coordination shells. The distributions for the systems with restraints over the CR2-N1-C1-C2 dihedral angle are given in red and for the systems without restraints in black.

Based on the dihedral angle distributions (Figure 6), one could expect the melting entropy of [C2] to decrease instead of increase upon restraining the dihedral angle. This is because the dihedral angle in the crystal already presents a sharper distribution than in the liquid phase, suggesting a larger entropy change in the liquid phase upon dihedral angle restraining than in the solid phase. If the entropy change $\Delta\Delta S_{m,dih}$ is only related to the dihedral angle distribution, $\Delta\Delta S_{m,dih} = \Delta S_{dih,L} - \Delta S_{dih,C} = (S_{dih,RL} - S_{dih,L}) - (S_{dih,RS} - S_{dih,S})$ should be negative. However, $\Delta\Delta S_{m,dih}$ is not the only source of entropy change. Results in Figure 7 suggest that the intermolecular organization increases more in the solid phase than in the liquid phase. Therefore, the change in the melting entropy due to the dihedral angle restraint can be decomposed into two terms (Equation 13), $\Delta\Delta S_{m,dih}$ and $\Delta\Delta S_{m,other,}$ where the first is related directly to the dihedral angle distributions and the latter is due to all the other molecular arrangements in the system. $\Delta\Delta S_{m,other}$ cannot be directly computed without a complete sampling of the phase space. The structural analyses suggest it is positive since the increase in ordering in the crystal due to the restraint is more intense than in the liquid phase. The total $\Delta\Delta S_m$ is known to be positive, $\Delta\Delta S_{m,other}$ must be positive and overcomes the negative term $\Delta\Delta S_{m,dih}$.

$$\Delta\Delta S_m = \Delta S_{m,restr.} - \Delta S_{m,unrestr.} = \Delta\Delta S_{m,dihed} + \Delta\Delta S_{m,other} \quad \text{(Eq 13)}$$

For [C10], the changes in the radial distribution functions involving the ionic portions are even more pronounced than for [C2] (Figure 7), suggesting a larger $\Delta\Delta S_{m,other}$. The dihedral angle distributions for the two phases are similar in this compound (Figure 6), indicating a less negative $\Delta\Delta S_{m,dih}$. Therefore, a larger $\Delta\Delta S_m$ is observed for [C10] than [C2] upon the dihedral angle restraint. These qualitative observations agree with the changes in the melting entropy, as shown in Table 2.

As a summary for this section, the same dihedral angle displayed different effects on the melting point of the two homologous ILs. The changes in this dihedral angle flexibility affects the intermolecular arrangement, resulting in non-trivial enthalpy and entropy contributions to the melting process. In the next section, the effects of other dihedral angles in the [C10] cation alkyl chain will be discussed.

**Effect of the decyl group dihedral angles**

In this section the effects of restraints of the other dihedral angles in the decyl group of [C10] are discussed. Each of these dihedral angles will be represented by a number as shown in Figure 8a. For example, the dihedral angle N1-C1-C2-C3 (see Figure 1 for atom names) is numbered as 0, the dihedral angle C1-C2-C3-C4 is numbered as 1, etc.

In the absence of any restraint, typical distributions of alkanes are found for these decyl dihedral angles. They all show a global maximum at 180° corresponding to the *trans* conformation and symmetrical local maxima at *ca.* ± 70° corresponding to the *gauche* conformation (Figures S3 to S10 in Supporting Information). Even in the crystal phase, *gauche* populations have been observed and the *gauche* fraction tends to increase at higher temperatures. In the liquid phase, no significant change was observed in the distributions with increasing temperature except a small broadening. The *gauche* fraction in the crystal (Figure 8 b) displays an interesting alternating behavior. The even-numbered dihedral angles have smaller populations of *gauche* defects than the odd numbered ones. The effect of temperature is stronger over the odd-numbered dihedral angles, which presents a larger increase in the *gauche* population when increasing the crystal temperature between 300 and 425 K. Beyond this temperature, the temperature effect becomes small and similar for every dihedral angle.

With increasing temperature, one can notice that for the odd-numbered dihedral angles, the dihedral angle distributions in the crystal converge to those of the liquid phase while the long range ordering remains (Figures S3 to S10). This implies that a significant disordering takes place in the apolar region due to the increase of the less favorable conformation. This behavior is demonstrated for a portion of the [C10] crystal in Figure 8c, where the increase of gauche defects and less perfect packing of the decyl groups at higher temperature is observed. On the other hand, the even-numbered dihedral angles have smaller *gauche* fractions than the liquid phase at all the studied temperatures. The existence of conformational disordering in the crystal even before the melting temperature was already observed experimentally for other ILs.[14,15,19]

The last dihedral angle (dihedral 7) presents the largest fraction of *gauche* defects in the crystal. Even in the liquid phase, while the other dihedral angles have nearly the same *gauche* fraction, this dihedral angle has a larger *gauche* fraction (Figure S10). This is because the terminal dihedral angle is less restricted than the others. Due to this peculiarity, some deviations will be noticed for dihedral angle 7 in comparison to the other odd-numbered dihedral angles.

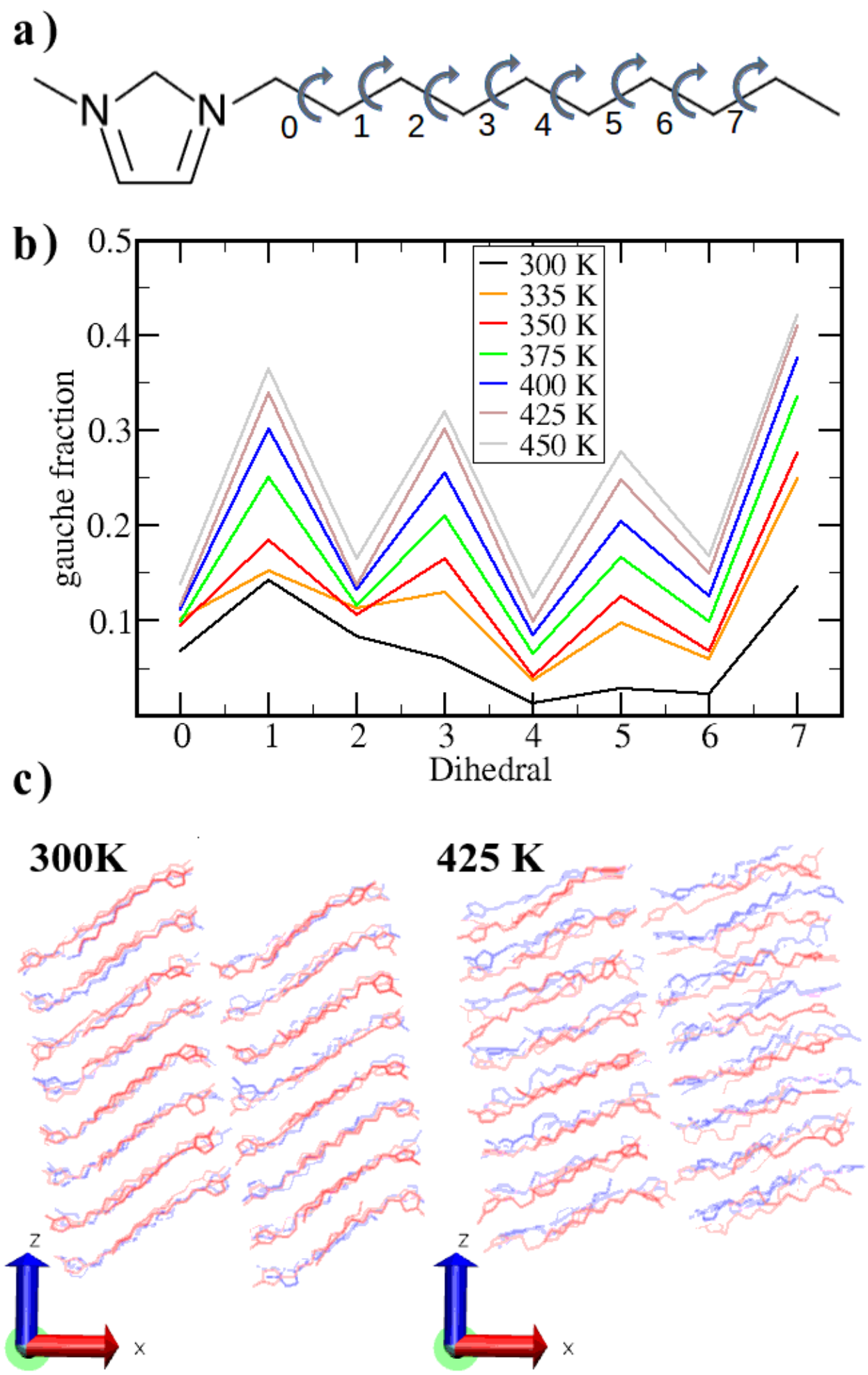

**Figure 8 a)** Numbering of the dihedral angles of the aliphatic tail of the [C10] cation, 0 is the N1-C1-C2-C3 dihedral and the next follow the sequence of carbon atoms in Figure 1. **b)** Fraction of *gauche* conformers for each dihedral angle in the solid phase at different temperatures. **c)** Representations of two layers of the [C10] crystal in the X direction at 300 K (left) and 425 K (right). For better visualization, only the nitrogen and carbon atoms are shown and the molecules were drawn in different colors depending on the position in the Y direction with red molecules closer and blue farther.

Due to the difference in distribution, it is expected that the flexibility of odd and even numbered dihedral angles affect the melting point in different ways. To study this effect, we applied the additional potential defined in Equation 11 to restrain one dihedral angle each time close to 180° (*trans* conformation). The computed free energy to restrain each dihedral angle is given in Figure 9 (top) for the crystal (black lines) and liquid (red lines) phases. The Helmholtz free energy to introduce the restraint is almost constant for every dihedral angle in the liquid phase except for dihedral angle 7 and to a smaller extent for dihedral angle 0. An alternating behavior is observed for the solid phase, which correlates with the *gauche* fraction shown in Figure 8b. The even-numbered dihedral angles that have small *gauche* fractions show small free energy changes when restrained to the *trans* conformation because they are already partially restrained due to intermolecular interactions. For a similar reason, the less restricted odd numbered dihedral angles experience higher free energy variations for the C→RC conversion.

The free energy difference between the restrained crystal and the restrained liquid were computed according to Equation 12. The results are shown in the Figure 9 (bottom) where a dashed line was included to represent the free energy difference between the liquid and the crystal in the absence of restraints. As expected, an alternating behavior emerges in an opposite way to that observed in $\Delta$A(C→RC). The even-numbered dihedral angles have large free energy differences because the liquid phase is more affected by the restraint than the solid phase, while the odd-numbered dihedral angles have small free energy changes because the two phases are affected to a similar extent.

It can also be seen in Figure 9 that, while the $\Delta G$(RC→RL) are nearly the same for every even-numbered dihedral angle restrained, $\Delta G$(RC→RL) increases for the odd-numbered dihedral angles when moving further away from the imidazolium ring. If the aliphatic chain is longer, it is likely that $\Delta G$(RC→RL) will converge to the same value at some distance from the imidazolium ring for every dihedral angle except the last one, which is an anomaly. This trend seems consistent with that in the *gauche* fraction shown in Figure 8.

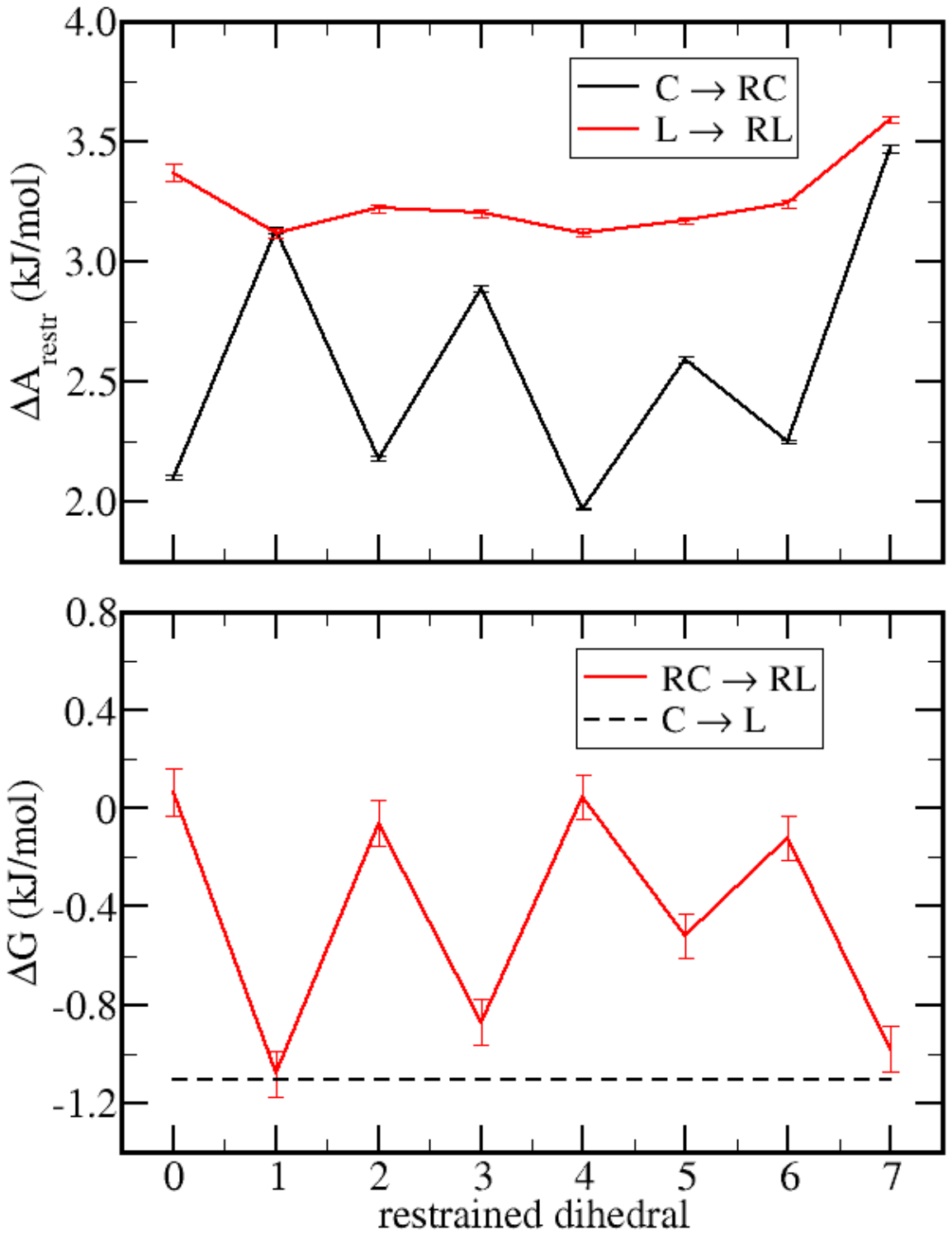

**Figure 9** Top: Calculated Helmholtz free energy change with restraint at each dihedral angle for the solid and the liquid of C10. Bottom: Gibbs free energy difference between the restrained liquid and the restrained crystal for each dihedral angle. The dotted line in the bottom panel shows the free energy difference between the liquid and the solid without restraints.

The alternating behavior observed in $\Delta G(\mathrm{RC}\rightarrow\mathrm{RL})$ (Figure 9) results in a similar odd-even alternating behavior in the calculated melting point (Figure 10 and Table 4). Therefore, the flexibility of different dihedral angles of the alkyl groups has different effects on the IL melting point. Restraining the dihedral angle that has a large number of defects in the solid phase does not result in appreciable changes in melting temperature, whereas restraining the ones that tend to have low fraction of defects in the crystal phase can increase the melting point by amounts as high as 20 K.

As opposed to the case of the CR2-N1-C1-C2 dihedral angle, in which both the melting enthalpy and the melting entropy changed appreciably with the restraints, for the other dihedral angles of the decyl group the changes in the melting enthalpy when compared to the unrestrained system are small and the changes in melting entropy are the dominant effect. Because the distribution of each dihedral angle is essentially the same in the liquid (except for dihedral angle 7), the reduction of the entropy in the liquid phase is similar regardless of which dihedral angle  restrained. However, in the solid phase the loss of entropy is smaller for the dihedral angles with smaller *gauche* populations, which implies a smaller $\Delta S_m$ for the systems with restraints over the even-numbered dihedral angles and therefore causes a larger increase in the melting point.

It is worth mentioning that although it does not have a significant effect on the melting point upon dihedral angle restraining, the melting enthalpy also exhibits an alternating behavior. As shown in Table 4, all the non-bonded and bonded contributions to enthalpy are affected by the dihedral angle restraints. Additional comments regarding these small but systematic variations of the energy components are given in Supporting Information file.

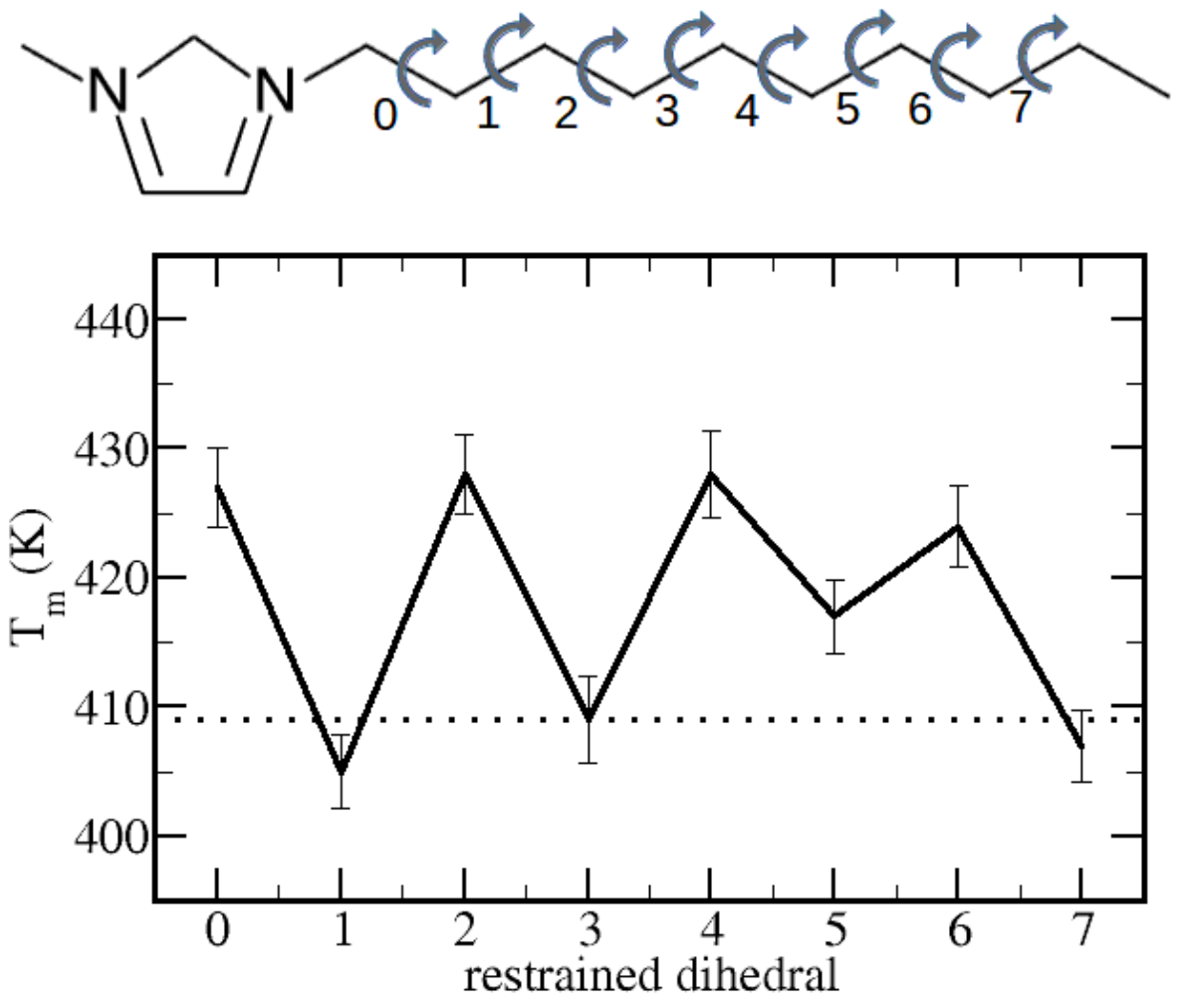

**Figure 10** Melting point of [C10] with each dihedral angle restrained. The horizontal dotted line shows the melting point in the absence of restraints.

The correlation between dihedral angles upon restraint were also studied. It was found that the restraint on one dihedral angle affects the *gauche* fraction of the other dihedral angles in the crystal phase (black lines in Figure 11) but not in the liquid phase (red lines in Figure 11). An alternating behavior in the correlation is also observed. The restraint on one dihedral angle does not affect appreciably the distribution of its immediate neighbor dihedral angles, but instead the *gauche* population of the next neighbor dihedral angles are suppressed. In other words, the presence of *gauche* defect for a given dihedral angle *n* will likely increase the probability of finding *gauche* defects for dihedral angles *n*-2 and *n*+2.

**Table 4** Melting points, melting enthalpies and entropies and differences in energy components between liquid and crystal at the melting point. Values are in K for melting point, kJ/mol for enthalpy and energy components and J $K^{-1}$ $mol^{-1}$ for the entropy. The subscripts represent the uncertainty in the last digit. Errors smaller than 0.005 are represented as zero.

| Dihedral | none | 0 | 1 | 2 | 3 | 4 | 5 | 6 | 7 |
|---|---|---|---|---|---|---|---|---|---|
| $T_m$ | $409_2$ | $427_3$ | $405_3$ | $424_3$ | $408_3$ | $426_3$ | $415_3$ | $423_3$ | $407_3$ |
| $\Delta H_m$ | $21.1_3$ | $20.7_1$ | $21.31_4$ | $20.5_1$ | $21.1_1$ | $20.7_1$ | $21.50_9$ | $20.59_9$ | $21.60_8$ |
| $\Delta S_m$ | $51.5_8$ | $48.5_1$ | $52.7_3$ | $48.4_2$ | $51.7_2$ | $48.4_1$ | $51.8_1$ | $48.7_1$ | $53.1_4$ |
| Electrost. | $10.61_4$ | $9.90_4$ | $10.39_3$ | $10.13_5$ | $9.71_5$ | $10.09_7$ | $10.48_5$ | $10.59_3$ | $10.51_2$ |
| VdW | $6.97_5$ | $7.93_8$ | $7.63_3$ | $7.56_8$ | $7.86_5$ | $7.66_6$ | $7.83_5$ | $7.12_5$ | $7.38_1$ |
| Bonds | $0.40_1$ | $0.38_0$ | $0.46_0$ | $0.35_0$ | $0.47_0$ | $0.39_0$ | $0.42_0$ | $0.37_0$ | $0.40_0$ |
| Angles | $1.99_4$ | $1.58_1$ | $1.95_1$ | $1.53_0$ | $2.04_1$ | $1.57_1$ | $1.89_1$ | $1.59_1$ | $2.09_2$ |
| Dihedrals | $1.02_2$ | $1.00_0$ | $0.97_1$ | $0.94_0$ | $1.09_1$ | $0.96_0$ | $1.05_1$ | $1.03_0$ | $1.24_2$ |

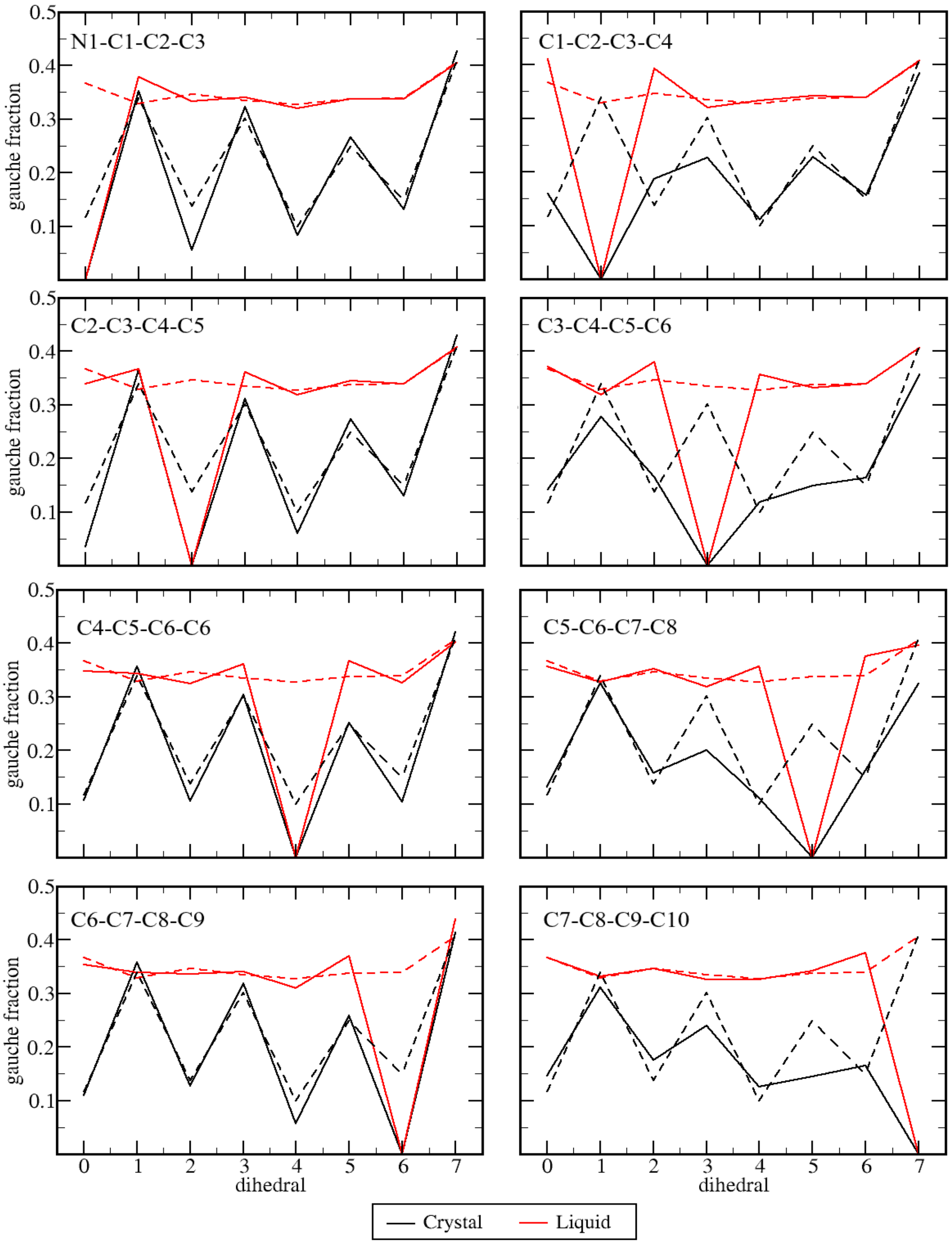

**Figure 11** Fraction of dihedral angles with *gauche* conformation for each restrained system (solid lines). The *gauche* fraction for the unrestrained system was included as a reference in each panel (dashed lines).

The odd-even alternating effect has been reported before and is well-known for linear alkanes[43] and its α and α,ω-derivates,[44,45] where the melting point as a function of carbon number, $n_C$, shows a zig-zag behavior instead of a monotonic shape. The increase in $T_m$ is larger when adding one $CH_2$ to a linear alkane with odd $n_C$ than that with even $n_C$. This effect is attributed to the more efficient packing in the crystal phase with even $n_C$. Despite the difficulty to synthesize IL with odd number of carbon atoms in the alkyl tail, examples of similar odd-even effects on the melting point[46,47] as well as on density in the liquid phase are known.[48] For both neutral molecules and ILs, the odd-even effect tends to disappear as the alkyl chain gets longer, consistent with our observation in melting points as shown in Figure 10. Such observation makes the tuning of the IL melting point by changing the alkyl chain length and flexibility complicated.

## CONCLUSIONS

The effects of dihedral angle flexibility on the melting point of alkyl-imidazolium-based ILs were studied using molecular dynamics simulations. Two ILs with different alkyl chain lengths in the cation, 1-ethyl-3-methylimidazolium hexafluorophosphate and 1-decyl-3-methylimidazolium hexafluorophosphate, were considered. The dihedral angle flexibility around the bond between the imidazolium ring and the alkyl group affects the relative packing and interaction between the charged portion of the cation and the anions, resulting in large variations in both the melting enthalpy and entropy. The resulting effect on the melting point is completely different for the two ILs. For the other dihedral angles of the decyl group, there is no significant effect over the ordering of the ionic part and the effect is mostly entropic, although some systematic enthalpic contributions were also noticed. An "odd-even" effect was found regarding these dihedral angles, which depends on how strong the rotational restraints imposed by the packing in the crystal are. The dihedral angles that are strongly restrained in the solid result in a large conformational entropy gain upon melting, thus, their flexibility contributes significantly to the melting point (up to 20 K). On the other hand, if a dihedral angle has considerable rotational freedom in the crystal, the flexibility affects both liquid and crystal phases in a similar way and no significant effect over the melting point is observed upon imposing a restraint.

If one can increase the dihedral angle rigidity or flexibility in the apolar chain of an IL by means of a chemical modification, for example, by introducing an ether group or a double bond, the effect on the $T_m$ will depend on the position in the chain. Any practical means to change the flexibility, however, will also change the intermolecular interactions, for instance, by introducing a local dipole by the substituent, which will make the situation complex. Nevertheless, our results can serve as an initial guide to select positions to make chemical modifications to control the melting point of the ionic liquid.

The thermodynamic cycle proposed here to study the effect of flexibility is computationally more efficient than the PSCP method. This new cycle also gives more information as the effect in each phase can be analyzed separately. The same strategy can be used to study other effects over the melting point once the melting free energy in a reference system is determined using the PSCP or other methods.

**Supporting Information**

See the Supporting Information for details about the reparametrization of the CR2-N1-C1-C2 dihedral angle, additional coments about the enthalpy effects of decyl group dihedral angles restraints, enthalpy variation with temperature, dihedral angle distributions at several temperatures, raw data used in the thermodynamic integration for introduction of restraints and distributions of N-C-C and C-C-C angles in the [C10] system.

**Data Availability**

The data that support the findings of this study are available from the corresponding author upon reasonable request.

**ACKNOWLEDGEMENTS**

KB and MCCR are indebted to FAPESP (grants 2017/12063-8, 2019/04785-9, and 2016/21070-5). MCCR is also indebted to CNPq (grant 301553/2017-3). YZ and EJM acknowledge support from the Air Force Office of Scientific Research under contract AFOSR FA9550-18-1-0321. We also thank the “Laboratório Nacional de Computação Científica (LNCC/MCTI, Brazil)” for the use of the supercomputer SDumont (https://sdumont.lncc.br).

**REFERENCES**

1. Dean, P. M.; Pringle, J. M.; MacFarlane, D. R. "Structural analyses of low melting organic salts: perspectives on ionic liquids" *Phys. Chem. Chem. Phys.* **12**, 9144-9153 (2010).
2. Weingärtner, H. "Understanding ionic liquids at the molecular level: facts, problems and controversies" *Angew. Chem. Int. Ed.* **47**, 654-670 (2008).
3. Zhou, F.; Liang, Y. M.; Liu, W. M. "Ionic liquid lubricants: designed chemistry for engineering applications" *Chem. Soc. Rev.* **38**, 2590-2599 (2009).
4. Krossing, I.; Slattery, J. M.; Daguenet, C.; Dyson, P. J.; Oleinikova, A.; Weingärtner, H. "Why are ionic liquids liquid? A simple explanation based on lattice and solvation energies" *J. Am. Chem. Soc.* **128**, 13427-13434 (2006).
5. Zahn, S.; Bruns, G.; Thar, J.; Kirchner, B. " What keeps ionic liquids in flow?" *Phys. Chem. Chem. Phys.* **10**, 6921-6924 (2008).
6. Hunt, P. A. "Why Does a Reduction in Hydrogen Bonding Lead to an Increase in Viscosity for the 1-Butyl-2,3-dimethyl-imidazolium-Based Ionic Liquids?" *J. Phys. Chem. B* **111**, 4844-4853 (2007).
7. Lobo Ferreira, A. I. M. C.; Rodrigues, A.; Villas, M.; Tojo, E.; Rebelo, L. P. N.; Santos, L. M. N. B. F. "On the Crystallization and Glass-Forming Ability of Ionic Liquids: Novel Insights into their Thermal Behavior" *ACS Sustainable Chem. Eng.* **7**, 2989-2997 (2019).
8. Serra, P. B. P.; Ribeiro, F. M. S.; Rocha, M. A. A.; Fulem, M.; Růžička, K.; Coutinho, J. A. P.; Santos, L. M. N. B. F. "Solid-liquid equilibrium and heat capacity trend in the alkylimidazolium PF6 series." *J. Mol. Liquids* **248**, 678-687 (2017).
9. Zhang, Y; Maginn, E. J. "Molecular dynamics study of the effect of alkyl chain length on melting points of [CnMIM][PF6] ionic liquids". *Phys. Chem. Chem. Phys.* **16**, 13489–13499 (2014).
10. Bernardino, K.; Lima, T. A.; Ribeiro, M. C. C. "Low-Temperature Phase Transitions of the Ionic Liquid 1-Ethyl-3-methylimidazolium Dicyanamide" *J. Phys. Chem. B* **123**, 9418–9427 (2019).
11. Jayaraman, S.; Maginn, E. J. "Computing the melting point and thermodynamic stability of the orthorhombic and monoclinic crystalline polymorphs of the ionic liquid 1-n-butyl-3-methylimidazolium chloride". *J. Chem. Phys.* **127**, 214504 (2007).
12. Holbrey, J. D.; Reichert, W. M.; Nieuwenhuyzen, M.; Johnson,S.; Seddon, K. R.; Rogers, R. D. "Crystal Polymorphism in 1-Butyl-3-Methylimidazolium Halides: Supporting Ionic Liquid Formation by Inhibition of crystallization." *Chem. Commun.* **14**, 1636−1637 (2003).

13. Ozawa, R.; Hayashi, S.; Saha, S.; Kobayashi, A.; Hamaguchi, H. “Rotational Isomerism and Structure of the 1-Butyl-3-Methylimidazolium Cation in the Ionic Liquid State.” *Chem. Lett.* **32**, 948−949 (2003).

14. Nishikawa, K.; Wang, S.; Katayanagi, H.; Hayashi, S.; Hamaguchi, H.; Koga, Y.; Tozaki, K. J. “Melting and Freezing Behaviors of Prototype Ionic Liquids, 1-Butyl-3-methylimidazolium Bromide and Its Chloride, Studied by Using a Nano-Watt Differential Scanning Calorimete” *J. Phys. Chem. B* **111**, 4894-4900 (2007).

15. Okajima, H.; Hamaguchi, H. “Unusually Long trans/gauche Conformational Equilibration Time during the Melting Process of BmimCl, a Prototype Ionic Liquid.” *Chem. Lett.* **40** 1308-1309 (2011).

16. Kotov, N.; Šturcová, A.; Zhigunov, A.; Raus, V.; Dybal, J. “Structural Transitions of 1-Butyl-3-methylimidazolium Chloride/Water Mixtures Studied by Raman and FTIR Spectroscopy and WAXS” *Cryst. Growth Des.* **16**, 1958-1967 (2016).

17. Saouane, S. ; Norman, S. E.; Hardacre, C.; Fabbiani, F. P. A. “Pinning down the solid-state polymorphism of the ionic liquid [bmim][PF6]” *Chem. Sci.* **4**, 1270-1280 (2013).

18. Faria, L. F. O.; Penna, T. C.; Ribeiro, M. C. C. “Raman Spectroscopic Study of Temperature and Pressure Effects on the Ionic Liquid Propylammonium Nitrate” *J. Phys. Chem. B* **117** 10905-10912 (2013).

19. Lima, T. A.; Paschoal, V. H.; Faria, L. F. O.; Ribeiro M. C. C. “Unraveling the Stepwise Melting of an Ionic Liquid” *J. Phys. Chem. B* **121**, 4650-4655 (2017).

20. Henderson, W. A.; Herstedt, M.; Young Jr. V. G., Passerini, S.; De Long, H. C.; Trulove, P. C. “Plastic Phase Transitions in N-Ethyl-N-methylpyrrolidinium Bis(trifluoromethanesulfonyl)imide” *Inorg. Chem.* **45**, 1412–1414 (2006).

21. Bernardino, K.; Goloviznina, K.; Gomes, M. C.; Pádua, A. A. H.; Ribeiro, M. C. C. “Ion pair free energy surface as a probe of ionic liquid structure” *J. Chem. Phys.* **152**, 014103 (2020).

22. Monteiro, M. J.; Camilo, F. F.; Ribeiro, M. C. C.; Torresi, R. M. “Ether-Bond-Containing Ionic Liquids and the Relevance of the Ether Bond Position to Transport Properties” *J. Phys. Chem. B.* **114**, 12488-12494 (2010).

23. Zhang, J.; Fang, S.; Qu, L.; Jin, Y.; Yang, L.; Hirano, S. “Synthesis, Characterization, and Properties of Ether-Functionalized 1,3-Dialkylimidazolium Ionic Liquids” *Ind. Eng. Chem. Res.* **53**, 16633-16643 (2014).

24. Terasawa, N.; Tsuzuki, S.; Umecky, T.; Saito, Y.; Matsumoto, H. “Alkoxy chains in ionic liquid anions; effect of introducing ether oxygen into perfluoroalkylborate on physical and thermal properties” *Chem. Commun.* **46**, 1730-1732 (2010).

25. Holbrey, J. D.; Reichert, W. M.; Rogers, R. D. "Crystal structures of imidazolium bis(trifluoromethanesulfonyl)imide 'ionic liquid' salts: the first organic salt with a cis-TFSI anion conformation" *Dalton Trans.* **15**, 2267-2271 (2004).

26. de Moura, A. F.; Bernardino, K.; de Oliveira, O. V.; Freitas, L. C. G. "Solvation of Sodium Octanoate Micelles in Concentrated Urea Solution Studied by Means of Molecular Dynamics Simulations" *J. Phys. Chem. B* **115**, 14582-14590 (2011).

27. Bernardino, K.; de Moura, A. F. "Aggregation Thermodynamics of Sodium Octanoate Micelles Studied by Means of Molecular Dynamics Simulations" *J. Phys. Chem. B* **117**, 7324-7334 (2013).

28. Eike, D. M; Brennecke, J. F.; Maginn, E. J. "Toward a robust and general molecular simulation method for computing solid-liquid coexistence" *J. Chem. Phys.* **122**, 014115 (2005).

29. Zhang, Y.; Maginn, E. J. "A comparison of methods for melting point calculations using molecular dynamics simulations". *J. Chem. Phys.* **136**, 144116 (2012).

30. S. Plimpton "Fast Parallel Algorithms for Short-Range Molecular Dynamics" *J. Comput. Phys.* **117**, 1–19 (1995).

31. Wang, J.; Wolf, R. M.; Caldwell, J. W.; Kollman, P. A. and Case, D. A. "Development and testing of a general amber force field" *J. Comput. Chem.* **25**, 1157–1174 (2004).

32. Zhang, Y; Maginn, E. J. "The effect of C2 substitution on melting point and liquid phase dynamics of imidazolium based-ionic liquids: insights from molecular dynamics simulations". *Phys. Chem. Chem. Phys.* **14**, 12157–12164 (2012).

33. F. Weigend and R. Ahlrichs "Balanced basis sets of split valence, triple zeta valence and quadruple zeta valence quality for H to Rn: Design and assessment of accuracy". *Phys. Chem. Chem. Phys.* **7**, 3297 (2005).

34. Neese, F. The ORCA program system, Wiley Interdiscip. Rev.: Comput. Mol. Sci. 2, 73-78 (2012).

35. Reichert, W. M.; Holbrey, J. D.; Swatloski, R. P.; Gutowski, K. E.; Visser, A. E.; Nieuwenhuyzen, M.; Seddon, K. R.; Rogers, R. D. "Solid-State Analysis of Low-Melting 1,3-Dialkylimidazolium Hexafluorophosphate Salts (Ionic Liquids) by Combined X-ray Crystallographic and Computational Analyses" *Crystal Growth & Design* **7**, 1106-1114 (2007).

36. Nosé, S. "A unified formulation of the constant temperature molecular-dynamics methods". *J. Chem. Phys.* **81**, 511–519 (1984).

37. M. Brehm and B. Kirchner "TRAVIS - A free Analyzer and Visualizer for Monte Carlo and Molecular Dynamics Trajectories". *J. Chem. Inf. Model.* **51** (8), 2007-2023 (2011).

*38.* Humphrey, W., Dalke, A. and Schulten, K., "VMD – Visual Molecular Dynamics". *J. Molec. Graphics* **14.1**, 33-38 (1996).

*39.* Grochola, G. "Constrained fluid λ-integration: Constructing a reversible thermodynamic path between the solid and liquid state". *J. Chem. Phys.* **120**, 2122 (2004).

*40.* Strehmel, V.; Laschewsky, A.; Krudelt, H.; Wetzel, H.; Görnitz, E., 2005. *Free Radical Polymerization of Methacrylates in Ionic Liquids.* In: *C. Brazel and R. Rogers, ed., Ionic Liquids in Polymer Systems, 1st ed. [online] American Chemical Society, pp.20,21. Available at: <https://pubs.acs.org/doi/book/10.1021/bk-2005-0913>*

*41.* Nemoto, F.; Kofu, M.; Yamamuro, O. "Thermal and structural studies of imidazolium-based ionic liquids with and without liquid-crystalline phases: the origin of nano-structure". *J. Phys. Chem. B* **119**, 5028–5034 (2015).

*42.* Domanska, U. and Marciniak, A. "Solubility of 1-Alkyl-3-methylimidazolium Hexafluorophosphate in Hydrocarbons". *J. Chem. Eng. Data* **48**, 451–456 (2003).

*43.* Yang, K.; Cai, Z.; Jaiswal, A.; Tyagi, M.; Moore, J. S.; Zhang, Y. "Dynamic odd-even effect in liquid n-alkanes near their melting points" *Angew. Chem. Int. Ed.* **55**, 14090-14095 (2016).

*44.* Badea, E.; Gatta, G. D.; D'Angelo, D.; Brunetti, B.; Rečková, Z. "Odd–even effect in melting properties of 12 alkane-α,ω-diamides" *J. Chem. Thermodyn.* **38**, 1546-1552 (2006).

*45.* Morishige, K. ; Kato, T. "Chain-length dependence of melting of n-alcohol monolayers adsorbed on graphite: n-hexanol, n-heptanol, n-octanol, and n-nonanol" *J. Chem. Phys.* **111**, 7095-7102 (1999).

*46.* Esperança, J. M. S. S.; Tariq, M.; Pereiro, A. B.; Araújo, J. M. M.; Seddon, K. R.; Rebelo, L. P. N. "Anomalous and Not-So-Common Behavior in Common Ionic Liquids and Ionic Liquid-Containing Systems" *Front. Chem.* **38**, 1546 – 1552 (2019).

*47.* Adamová, G.; Gardas, R. L.; Rebelo, L. P. N.; Robertson, A. J.; Seddon, K. R. "Alkyltrioctylphosphonium chloride ionic liquids: synthesis and physicochemical properties" *Dalton Trans.* **40**, 12750-12764 (2011).

*48.* Adamová, G.; Canongia Lopes, J. N.; Rebelo, L. P. N.; Santos, L. M. N. B.; Seddon, K. R.; Shimizu, K. "The alternation effect in ionic liquid homologous series" *Phys. Chem. Chem. Phys.* **16**, 4033-4038 (2014).

# Supporting Information for

# Effect of Alkyl-group Flexibility on the Melting Point of Imidazolium-based Ionic Liquids

Kalil Bernardino,[1*] Yong Zhang[2],

Mauro C. C. Ribeiro[1], Edward Maginn[2]

[1] *Laboratório de Espectroscopia Molecular, Departamento de Química Fundamental, Instituto de Química, Universidade de São Paulo, Av. Prof. Lineu Prestes 748, 05508-000, Brazil*

[2] *Department of Chemical and Biomolecular Engineering, University of Notre Dame, Notre Dame, Indiana, 46556, USA.*

***email: kalil.bernardino@gmail.com**

**Contents:**

# 1. Reparametrization of the CR2-N1-C1-C2 dihedral

Despite predicting the correct position for the global minimum for the torsion of the CR2-N1-C1-C2 dihedral angle (see Figure 1 in the manuscript for atom names), the GAFF parameters employed in the previous work (reference 7 of manuscript) of our group regarding the melting calculation of the 1-alkyl-3-metil-imidazolium perfluorborates do not describe properly the position of the secondary minimum expected to happen at 0º (which corresponds to the carbon C2 in the same plane defined by the imidazolium ring and in the same side of CR2 atom) and also underestimate the barrier at 180º, as can be seen by the comparison with *ab initio* MP2 calculations with Def2-TVZPPD basis set (Figure S1).

Both the force field (black curve) and the MP2 (red curve) calculations were performed by restraining the CR2-N1-C1-C2 dihedral angle while optimizing every other coordinate of the 1-ethyl-3-methyl-imidazolium cation. MP2 calculations for each dihedral value were performed with ORCA 4.0.1.2 and the force field optimizations were performed with LAMMPS. For the latter, a restrained potential was applied in order to keep the dihedral angle nearly fixed along the energy minimization and the effect of this restrained was subtracted from the final potential energy.

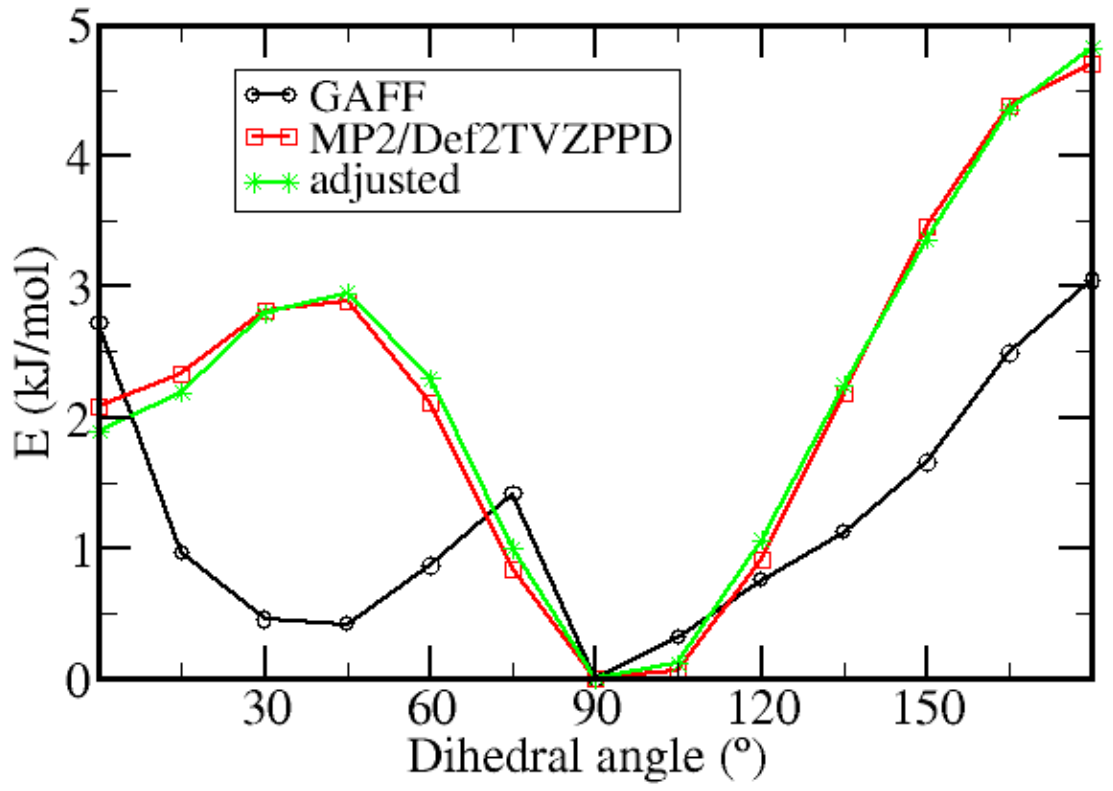

**Figure S1** Potential energy for each value of the dihedral angle CR2-N1-C1-C2 at [C2] cation with GAFF parameters, a MP2 calculation and with the adjusted dihedral potential. The energy of the global minimum was shifted to zero.

The GAFF force field employs the CHARMM type dihedrals, defined in Equation S1, where $n$ assumes integer values and $\phi$ is the dihedral angle, with the parameters $K_n$ and $d_n$ given in Tables S1 and S2, respectively. Notice that the potential energy in Figure S1 also include all other bonded and non-bonded contributions, as defined in the GAFF force field, and not only the torsion potential defined in Equation S1.

$$E_{dihed}(\phi)=\sum_{n=1}^{n=5} K_n\left[1+\cos\left(n\phi-d_n\right)\right] \quad \text{(Equation S1)}$$

The same functional form of Equation S1 was used to determine the new parameters by computing the difference between MP2 and GAFF potential energy surfaces and then subtracting the contribution that arrives directly from the original torsion potential of Equation S1. The $K_n$ parameters in Equation S1 were optimized in order to fit this difference while every $d_n$ was set to zero. The new parameters are given in Table S1 and Table S2 and the potential energy surface with the adjusted parameters is given as the green curve in Figure S1, which match almost exactly the results from the MP2 calculations.

**Table S1** GAFF parameters and adjusted $K_n$ parameters for the CR2-N1-C1-C2 dihedral (values in kcal/mol).

| Parameter | $K_1$ | $K_2$ | $K_3$ | $K_4$ | $K_5$ |
|---|---|---|---|---|---|
| GAFF | -0.05342 | 0.0797131 | 0.192789 | 0.0478 | 0 |
| Adjusted | 0.04389 | 0.01538 | -0.16753 | -0.24363 | -0.01803 |

**Table S2** GAFF parameters and adjusted $d_n$ parameters for the CR2-N1-C1-C2 dihedral (values in degree).

| Parameter | $d_1$ | $d_2$ | $d_3$ | $d_4$ | $d_5$ |
|---|---|---|---|---|---|
| GAFF | 0 | 180 | 0 | 180 | 0 |
| Adjusted | 0 | 0 | 0 | 0 | 0 |

## 2. Enthalpy variation with the temperature

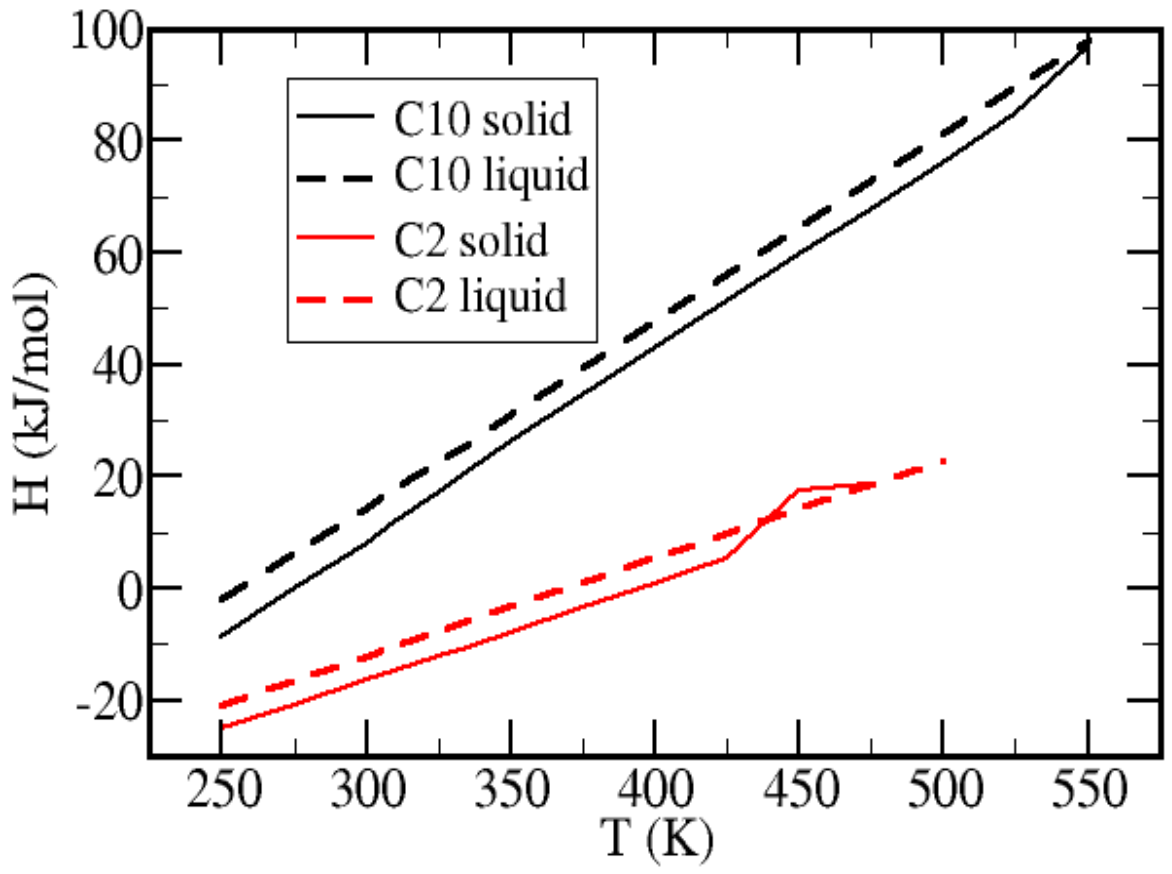

**Figure S2** Variation of the enthalpy of the solid (solid lines) and of the liquid (dashed lines) with the temperature for both C2 and C10. At high temperatures, a break occur at the curves for the solid due to spontaneous melting at 425 K for [C2] and at 525 K for [C10]. The region after the discontinuity due to the melting was not included in the linear regression used to obtain the parameters for the Gibbs-Helmholtz equation.

## 3. Dihedral angles distributions

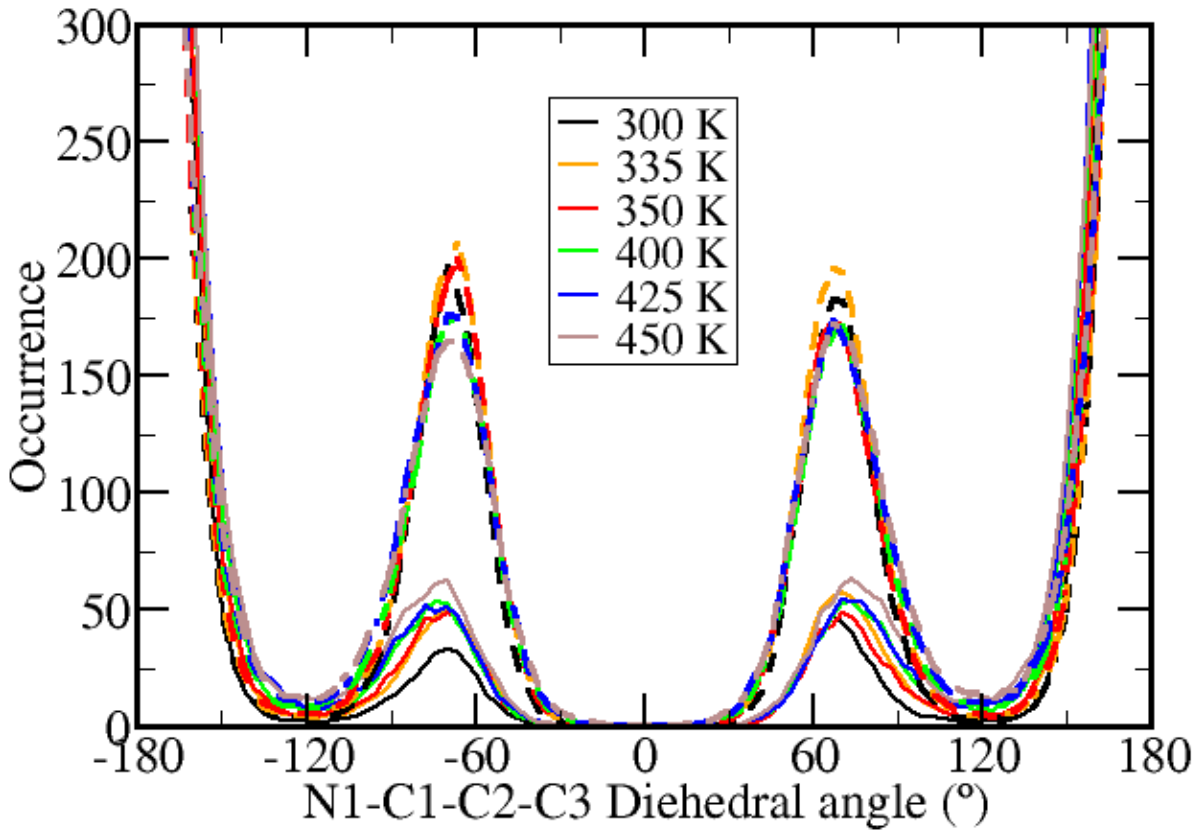

**Figure S3** Distribution of the N1-C1-C2-C3 dihedral angle (dihedral 0) for the unrestrained crystal (solid lines) and unrestrained liquid (dashed lines) at different temperatures.

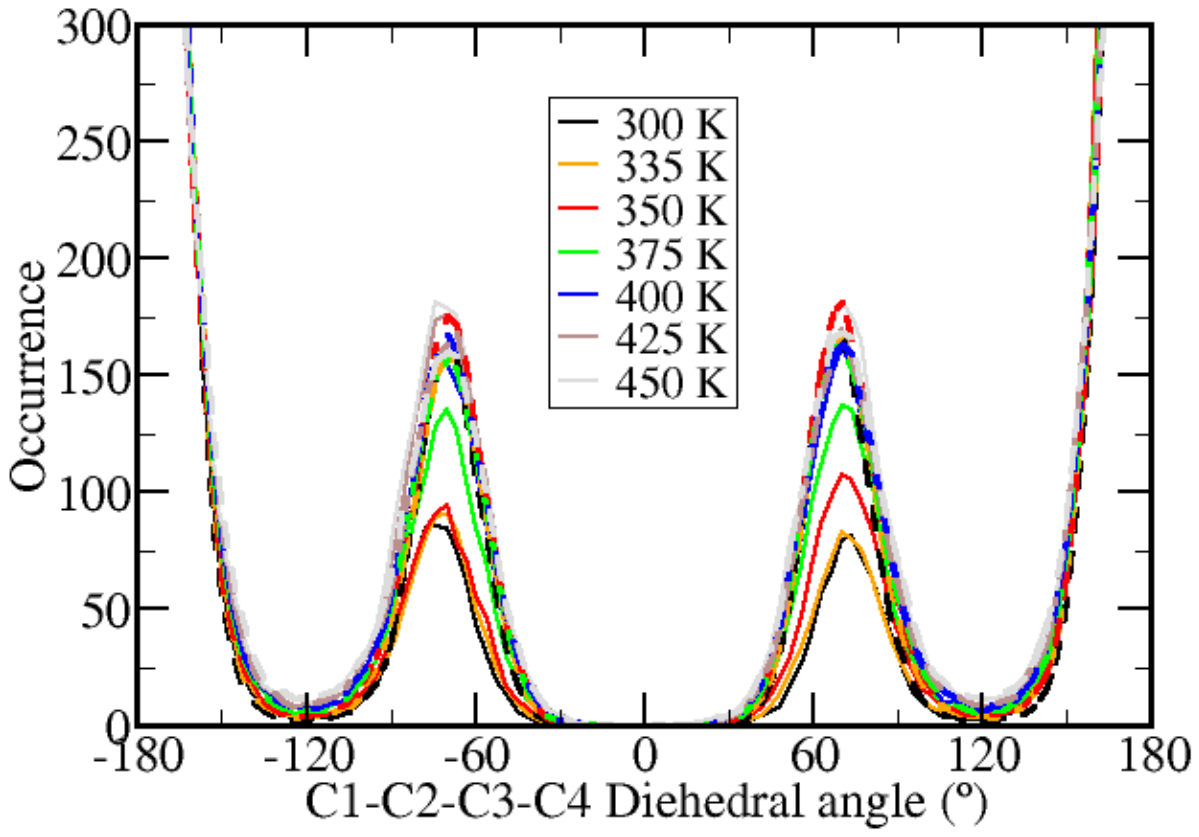

**Figure S4** Distribution of the C1-C2-C3-C4 dihedral angle (dihedral 1) for the unrestrained crystal (solid lines) and unrestrained liquid (dashed lines) at different temperatures.

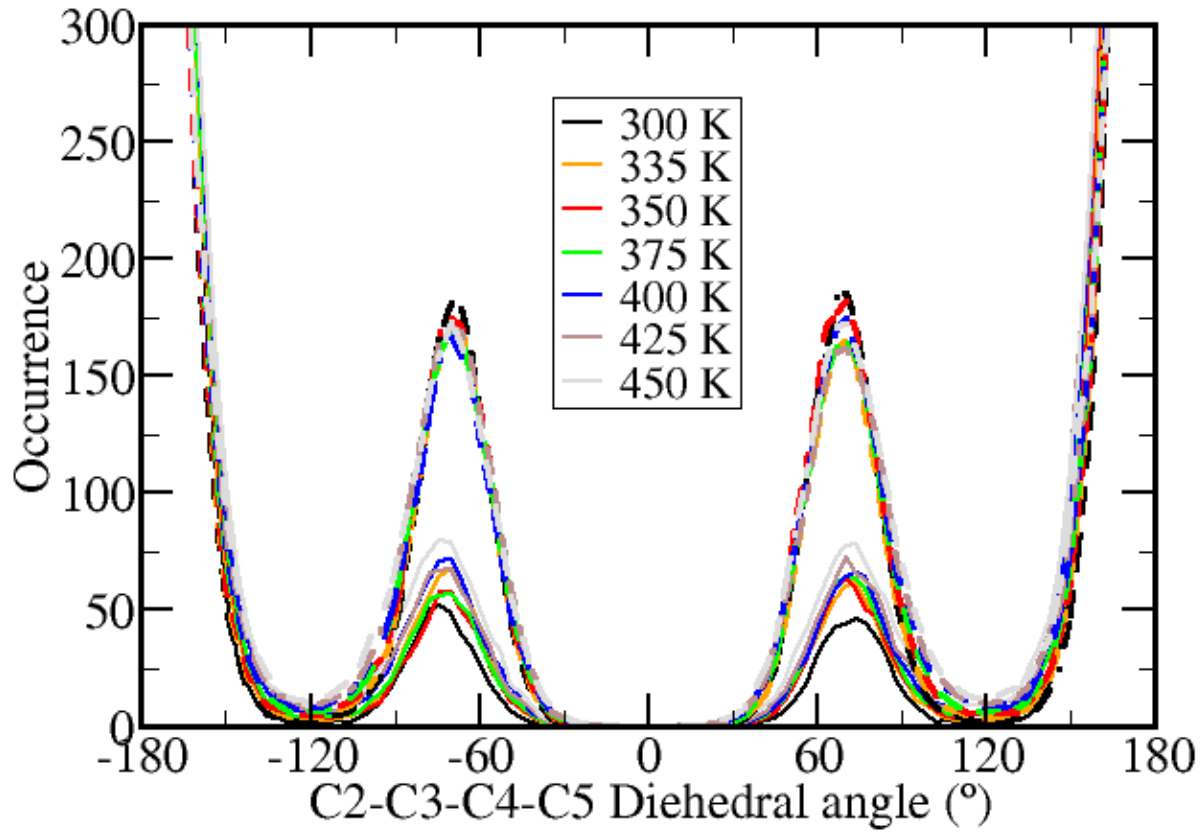

**Figure S5** Distribution of the C2-C3-C4-C5 dihedral angle (dihedral 2) for the unrestrained crystal (solid lines) and unrestrained liquid (dashed lines) at different temperatures.

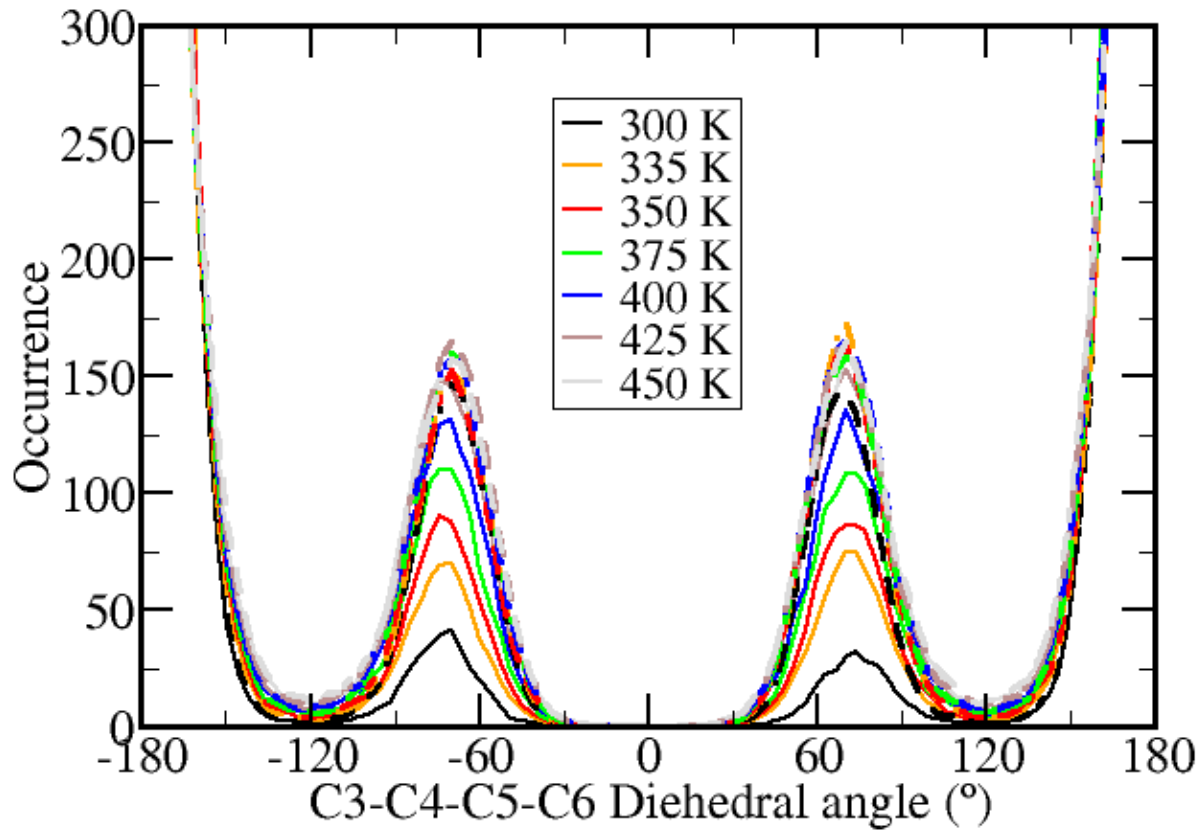

**Figure S6** Distribution of the C3-C4-C5-C6 dihedral angle (dihedral 3) for the unrestrained crystal (solid lines) and unrestrained liquid (dashed lines) at different temperatures.

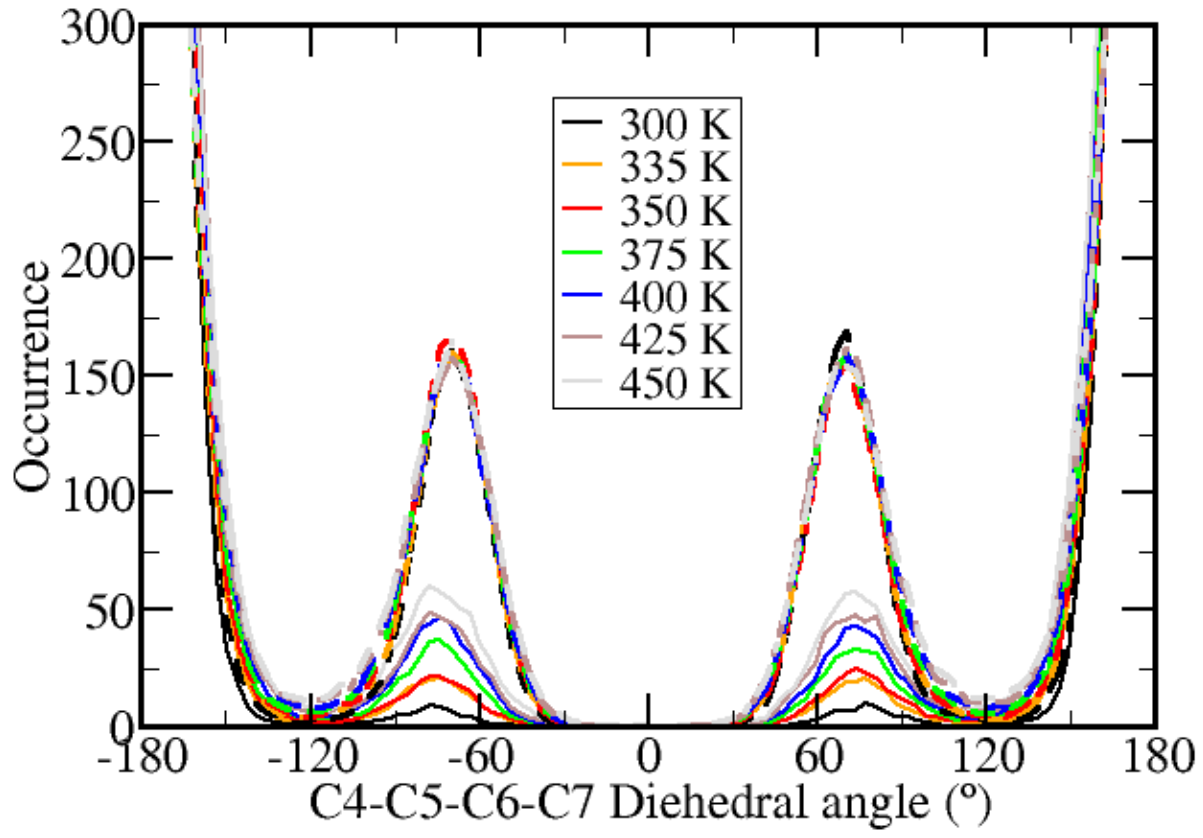

**Figure S7** Distribution of the C4-C5-C6-C7 dihedral angle (dihedral 4) for the unrestrained crystal (solid lines) and unrestrained liquid (dashed lines) at different temperatures.

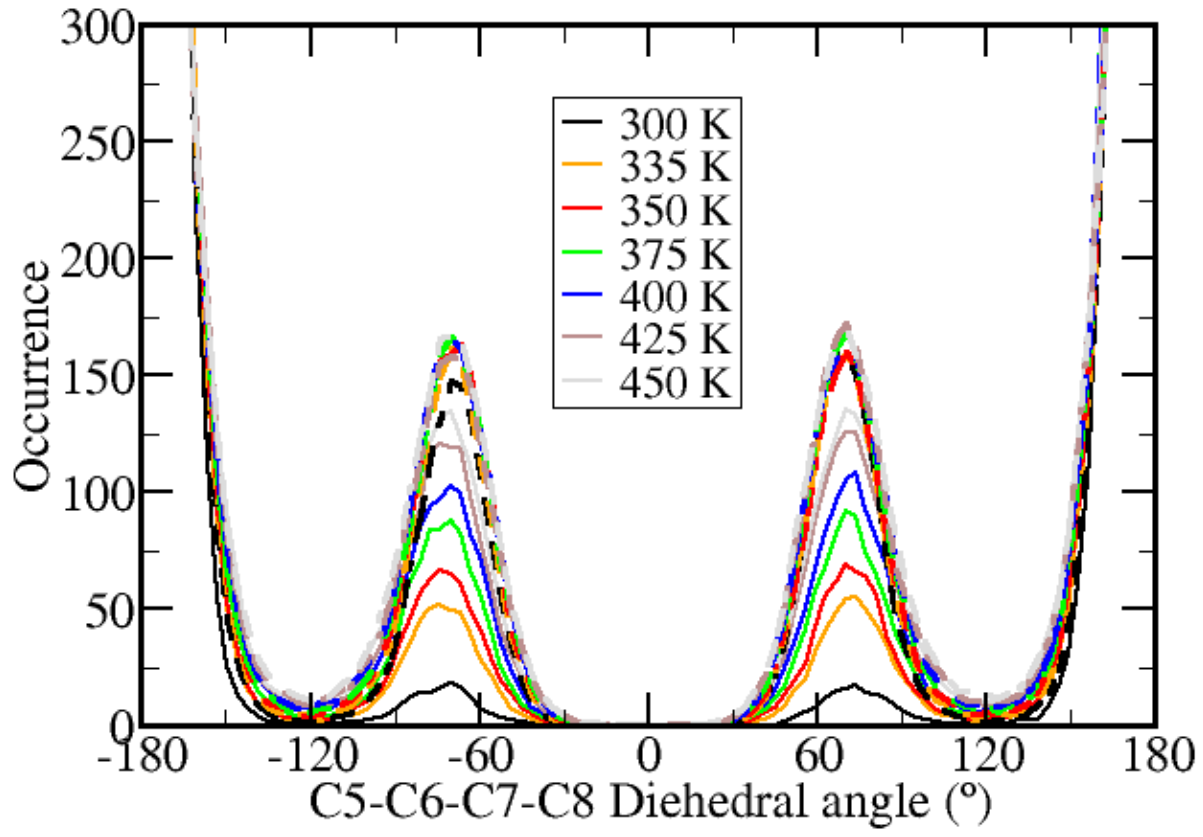

**Figure S8** Distribution of the C5-C6-C7-C8 dihedral angle (dihedral 5) for the unrestrained crystal (solid lines) and unrestrained liquid (dashed lines) at different temperatures.

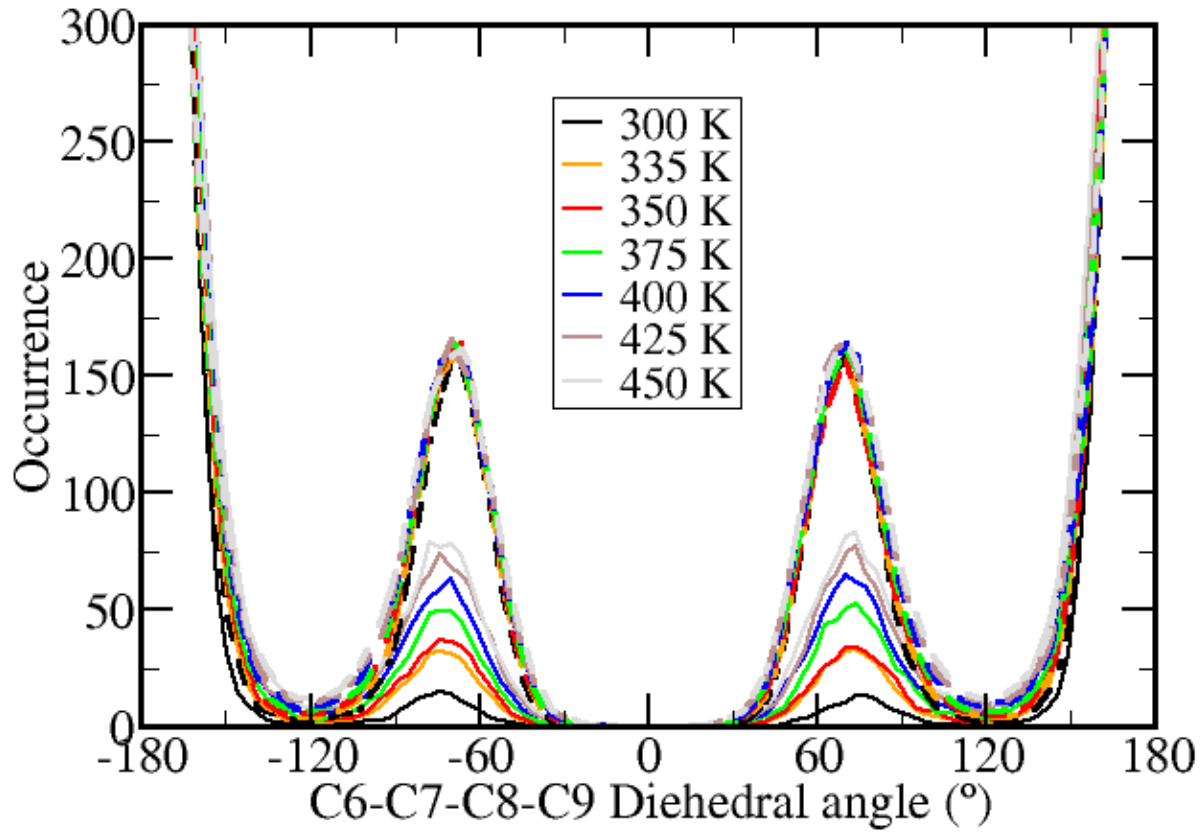

**Figure S9** Distribution of the C6-C7-C8-C9 dihedral angle for the unrestrained crystal (solid lines) and unrestrained liquid (dashed lines) at different temperatures.

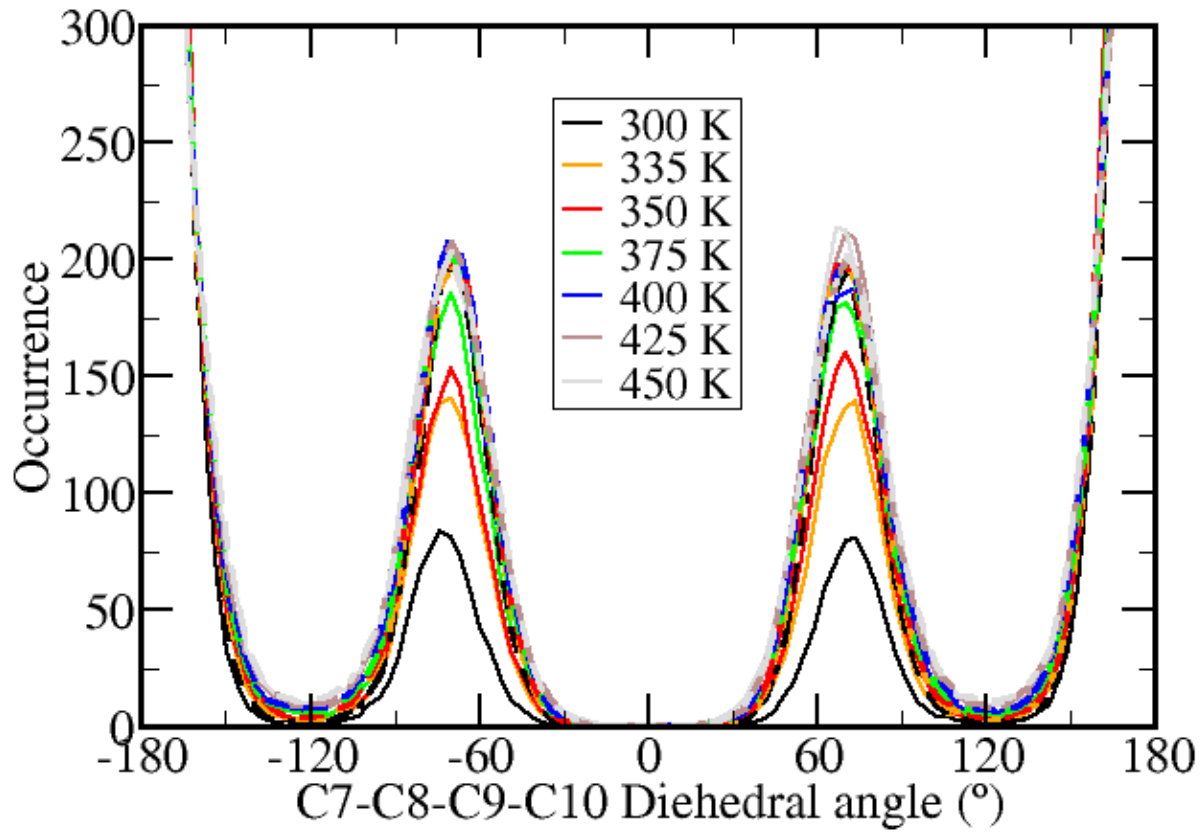

**Figure S10** Distribution of the C7-C8-C9-C10 dihedral angle (dihedral 7) for the unrestrained crystal (solid lines) and unrestrained liquid (dashed lines) at different temperatures.

# 4. Thermodynamic integration to introduce dihedrals restraints

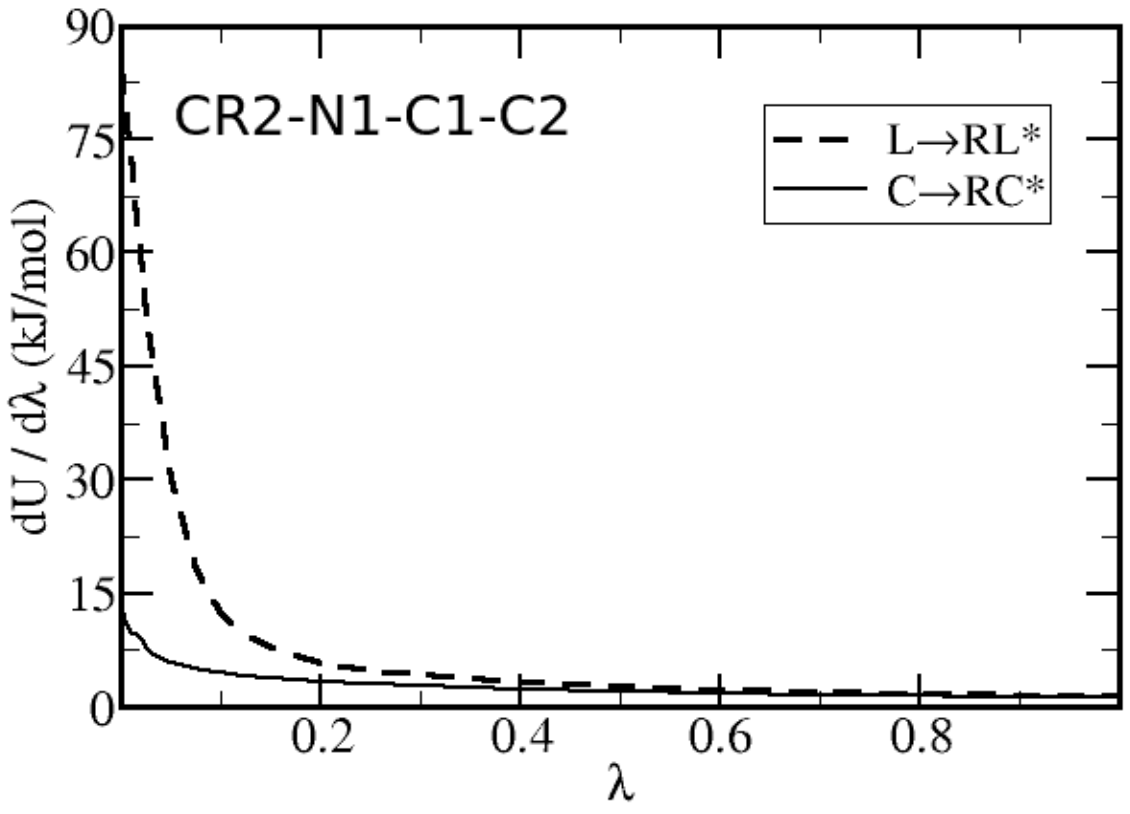

**Figure S11** Derivative of the system energy U with the coupling parameter λ for the introduction of the restraint (Equation 11) over the CR2-N1-C1-C2 dihedral angle in the liquid (L) and in the crystal (C) phases of [C2] at constant volume (steps a and c in Figure 3).

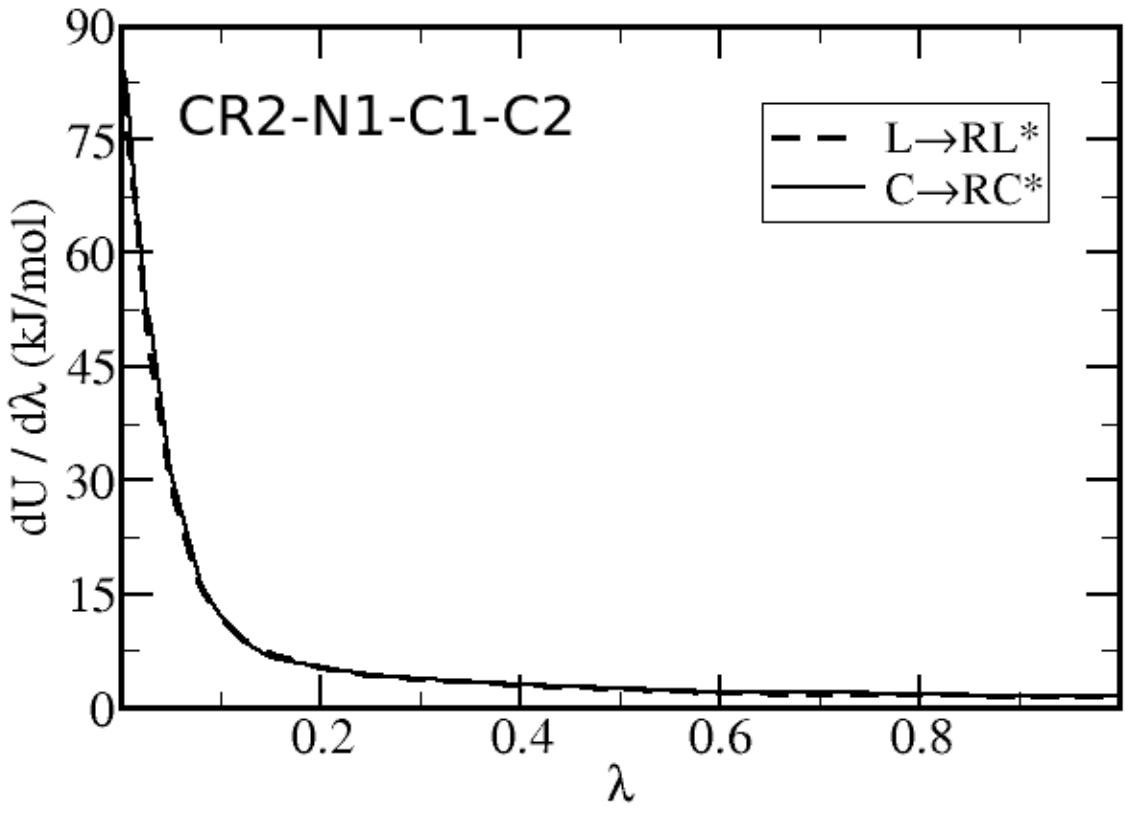

**Figure S12** Derivative of the system energy U with the coupling parameter λ for the introduction of the restraint (Equation 11) over the CR2-N1-C1-C2 dihedral angle in the liquid (L) and in the crystal (C) phases of [C10] at constant volume (steps a and c in Figure 3).

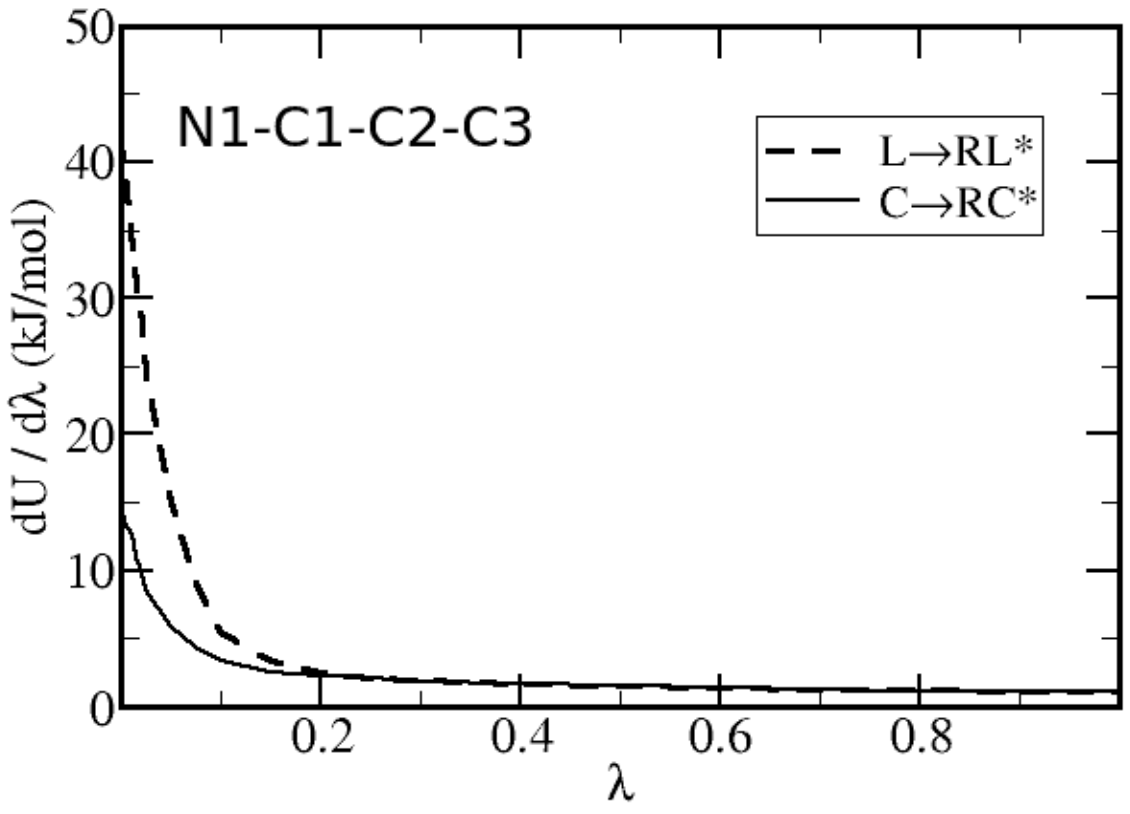

**Figure S13** Derivative of the system energy U with the coupling parameter λ for the introduction of the restraint (Equation 11) over the N1-C1-C2-C3 dihedral angle (dihedral 0) in the liquid (L) and in the crystal (C) phases of [C10] at constant volume (steps a and c in Figure 3).

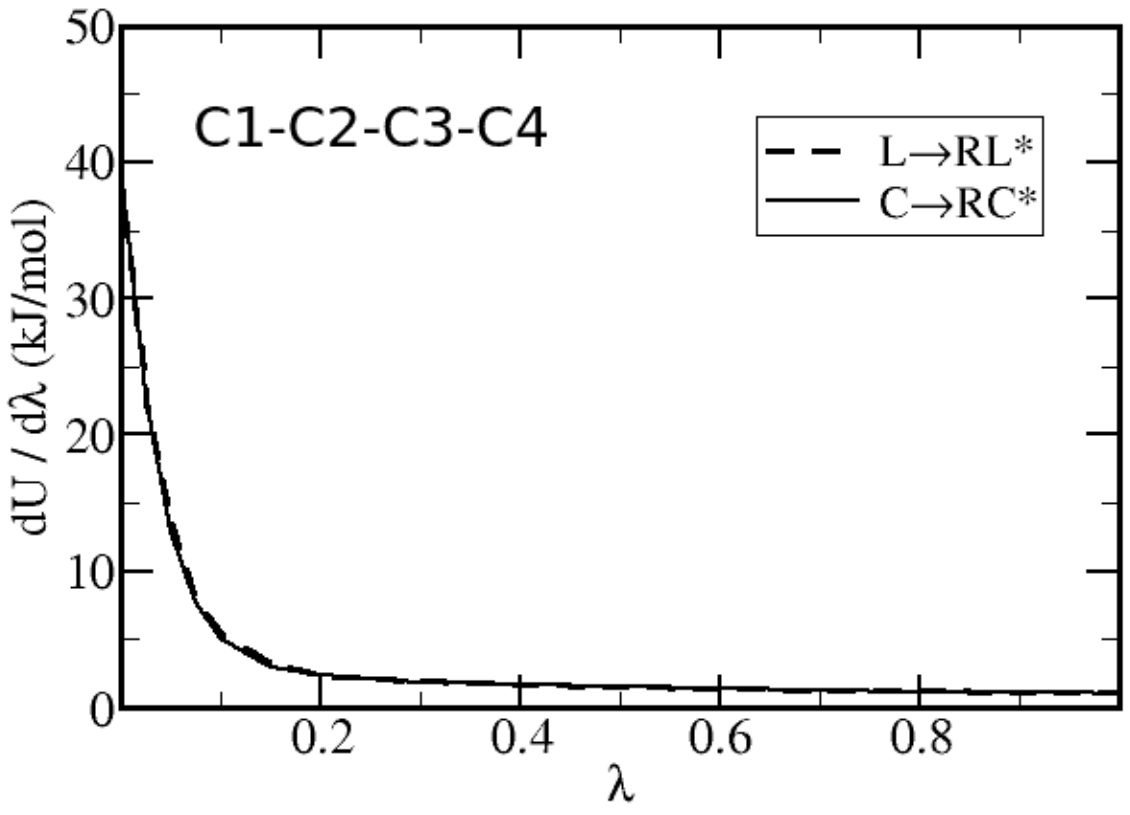

**Figure S14** Derivative of the system energy U with the coupling parameter λ for the introduction of the restraint (Equation 11) over the C1-C2-C3-C4 dihedral angle (dihedral 1) in the liquid (L) and in the crystal (C) phases of [C10] at constant volume (steps a and c in Figure 3).

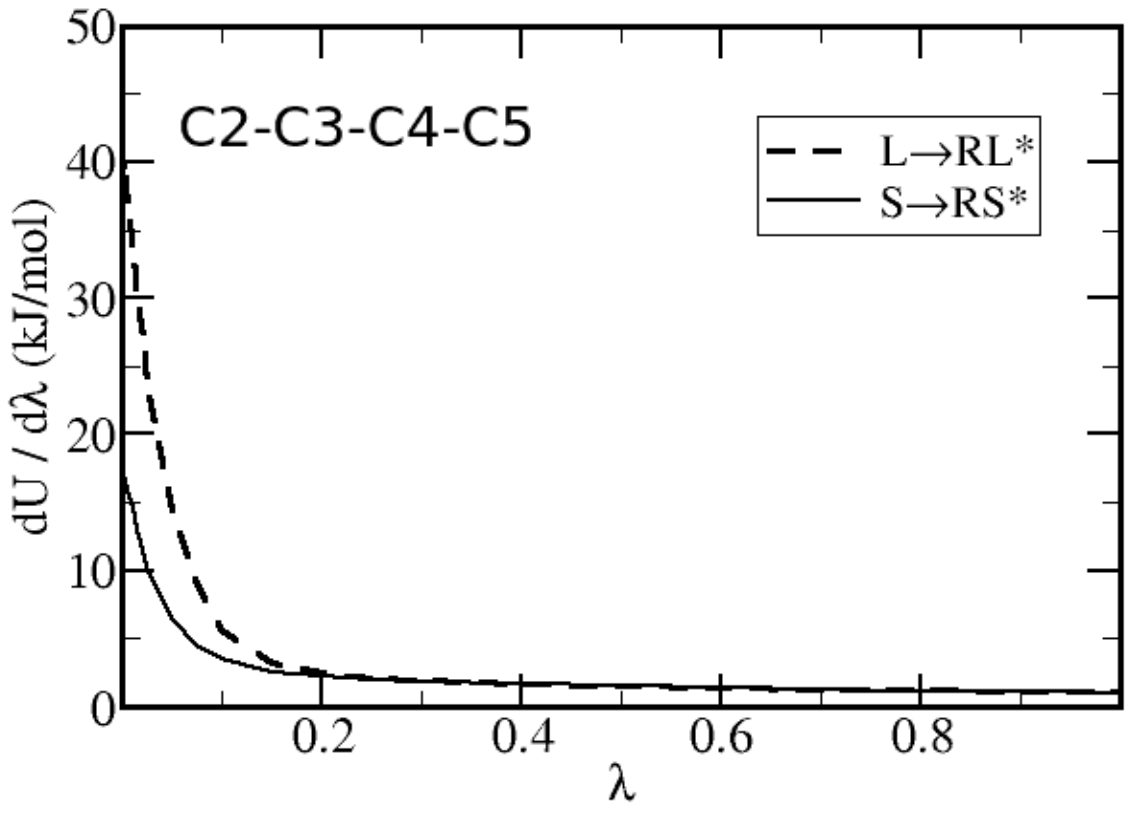

**Figure S15** Derivative of the system energy U with the coupling parameter λ for the introduction of the restraint (Equation 11) over the C2-C3-C4-C5 dihedral angle (dihedral 2) in the liquid (L) and in the crystal (C) phases of [C10] at constant volume (steps a and c in Figure 3).

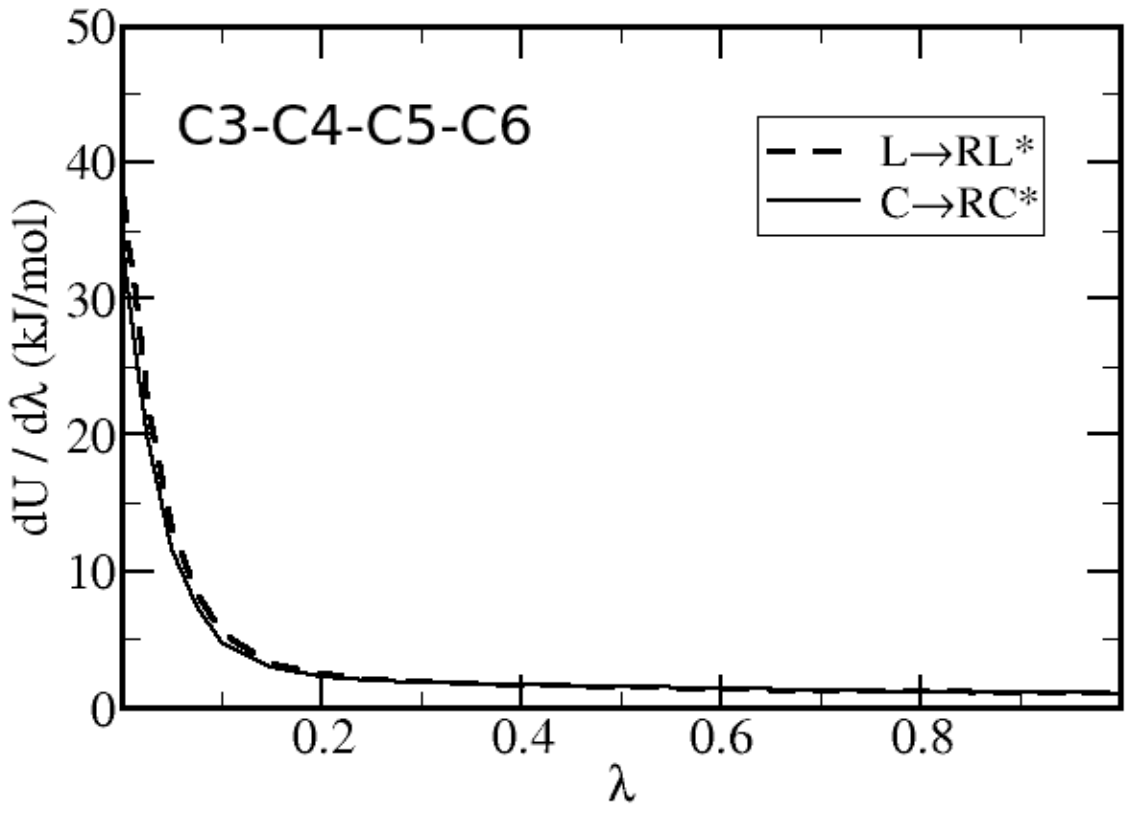

**Figure S16** Derivative of the system energy U with the coupling parameter λ for the introduction of the restraint (Equation 11) over the C3-C4-C5-C6 dihedral angle (dihedral 3) in the liquid (L) and in the crystal (C) phases of [C10] at constant volume (steps a and c in Figure 3).

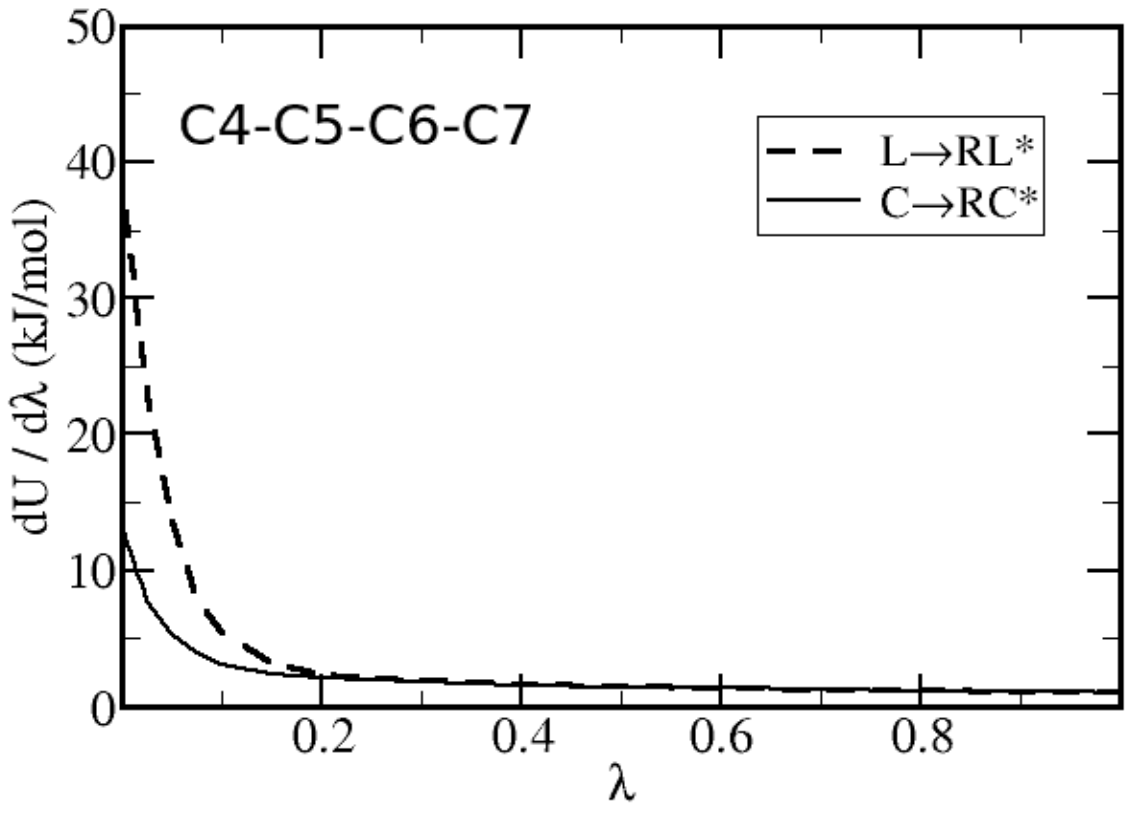

**Figure S17** Derivative of the system energy U with the coupling parameter λ for the introduction of the restraint (Equation 11) over the C4-C5-C6-C7 dihedral angle (dihedral 4) in the liquid (L) and in the crystal (C) phases of [C10] at constant volume (steps a and c in Figure 3).

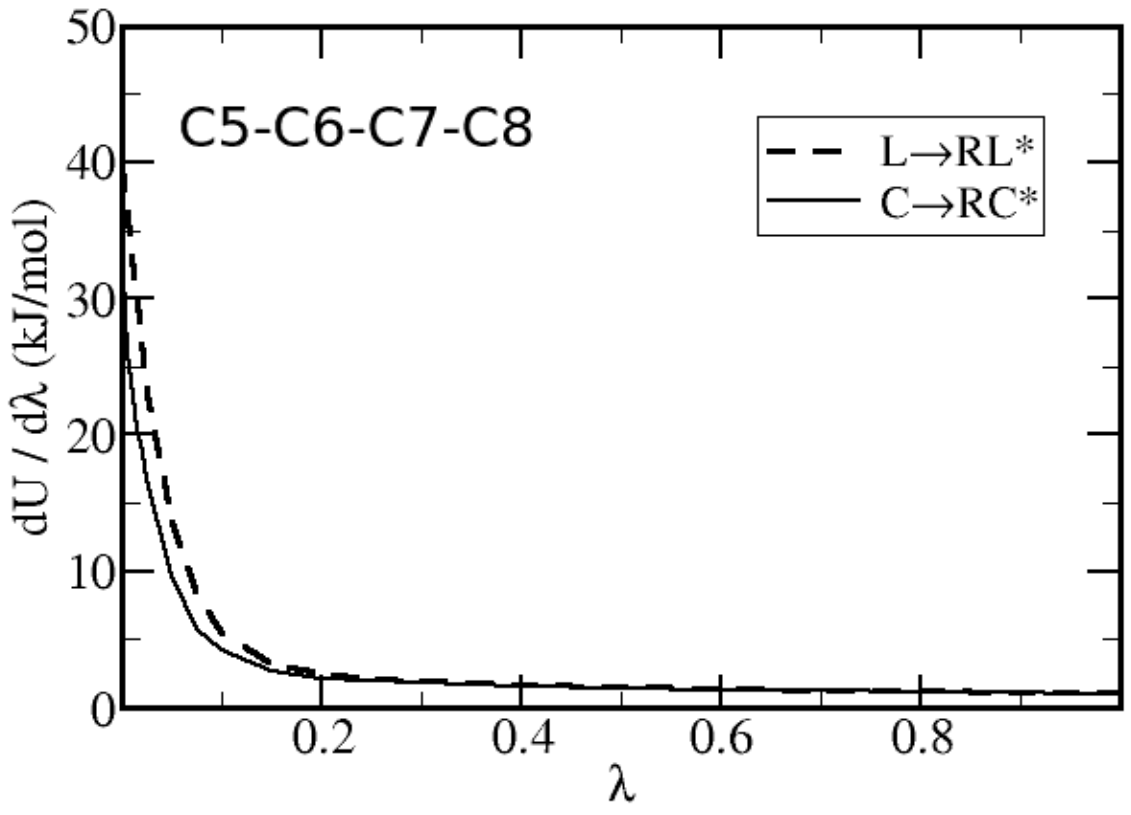

**Figure S18** Derivative of the system energy U with the coupling parameter λ for the introduction of the restraint (Equation 11) over the C5-C6-C7-C8 dihedral angle (dihedral 5) in the liquid (L) and in the crystal (C) phases of [C10] at constant volume (steps a and c in Figure 3).

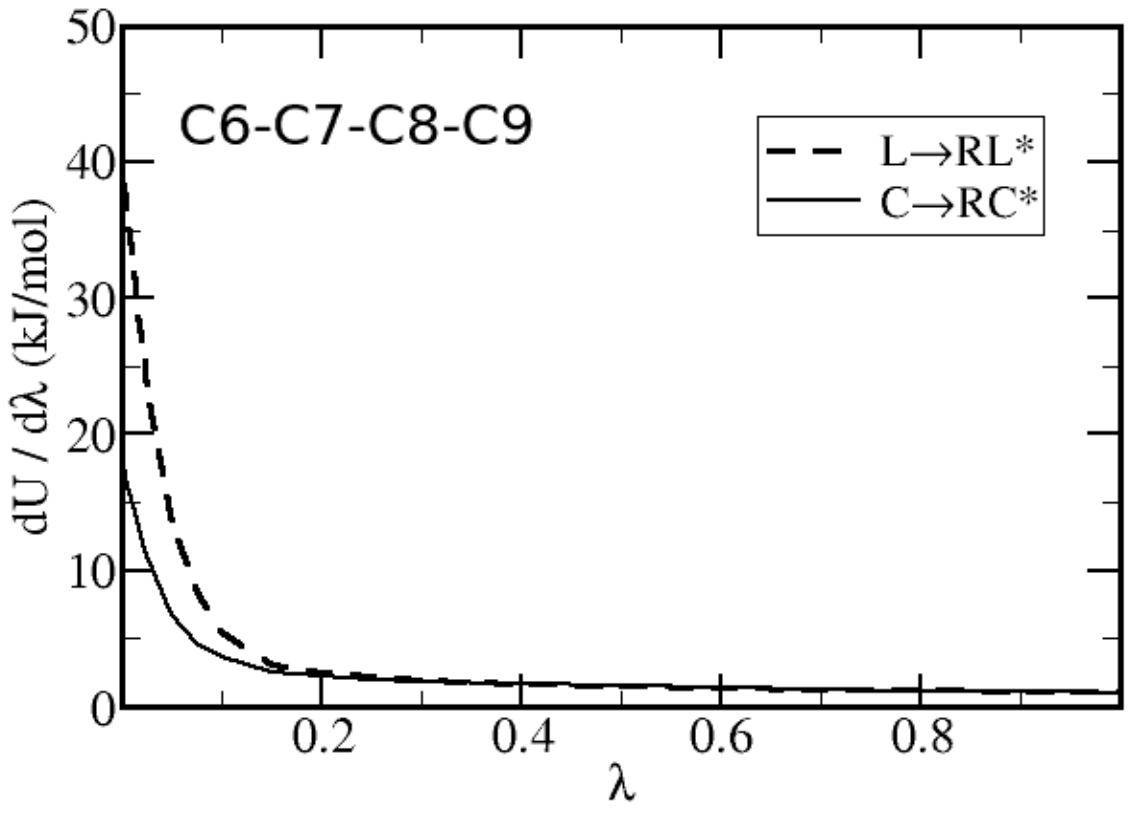

**Figure S19** Derivative of the system energy U with the coupling parameter λ for the introduction of the restraint (Equation 11) over the C6-C7-C8-C9 dihedral angle (dihedral 6) in the liquid (L) and in the crystal (C) phases of [C10] at constant volume (steps a and c in Figure 3).

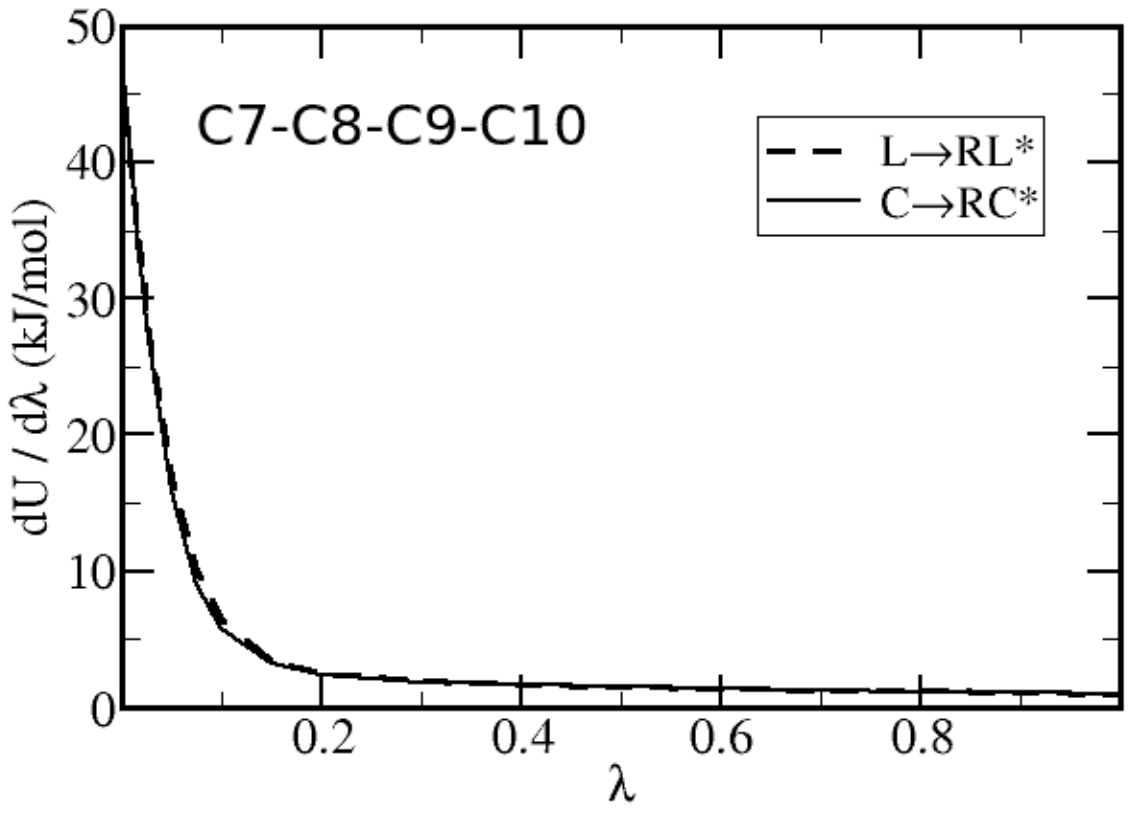

**Figure S20** Derivative of the system energy U with the coupling parameter λ for the introduction of the restraint (Equation 11) over the C7-C8-C9-C10 dihedral angle (dihedral 7) in the liquid (L) and in the crystal (C) phases of [C10] at constant volume (steps a and c in Figure 3).

## 5. Additional comments about the effects of dihedral restraints over melting enthalpy

Regarding the decyl group in the [C10] system, entropy changes upon restrains dominates the effect over the melting point but is partially offset by the change of the enthalpy, which also displays an alternating behavior related to the position of the restrained dihedral angle. The changes in the enthalpy can be further split into inter and intramolecular terms (Table 4), but the changes in each component are small when compared with the larger effects over both electrostatic and van der Waals terms in the case of the CR2-N1-C1-C2 dihedral (Table 2). No single component seems to dominate the differences observed in the melting enthalpy. However, some trends can be noticed, especially the fact that the van der Waals contribution increases in relation to the unrestrained system for every dihedral angle, indicating a better packing of the tails in the solid phase whenever a dihedral angle is forced to exist only in *trans* conformation.

There is a noticeable and unexpected alternating behavior regarding the angle potential component and a similar but smaller effect regarding the energy associated with bond stretching energy. Examining the distribution of each N-C-C and C-C-C angle in the solid and for the liquid phase as a function of dihedral angles restraints (Figures S21 to S29), one can see small shifts to smaller values in the distributions of the C-C-C angles involving the same carbon atoms of the dihedral angle over which the restraint is applied. For the liquid this effect is similar for the odd and the even dihedrals involving the atoms of the given angle. In the crystal, however, it is more significant for the odd-numbered dihedrals than for the even ones. The shift to small angle values brings the maximum of the distribution close to the position of the minimum of the angle potential defined in the force field (112.81º for N-C-C and 110.63º for C-C-C angles, vertical dotted lines in the Figures S21 to S29), thus reducing the average angle energy. These small angle shifts when restraining the dihedrals to the *trans* conformation are not an artifact of the force field. MP2/Def2-TVZPD geometry optimizations were performed for a butane molecule in both gauche and trans conformation and a reduction of 0.84º was observed for the C-C-C angles when going from *gauche* to *trans* conformation.

## 6. N-C-C and C-C-C angles distributions

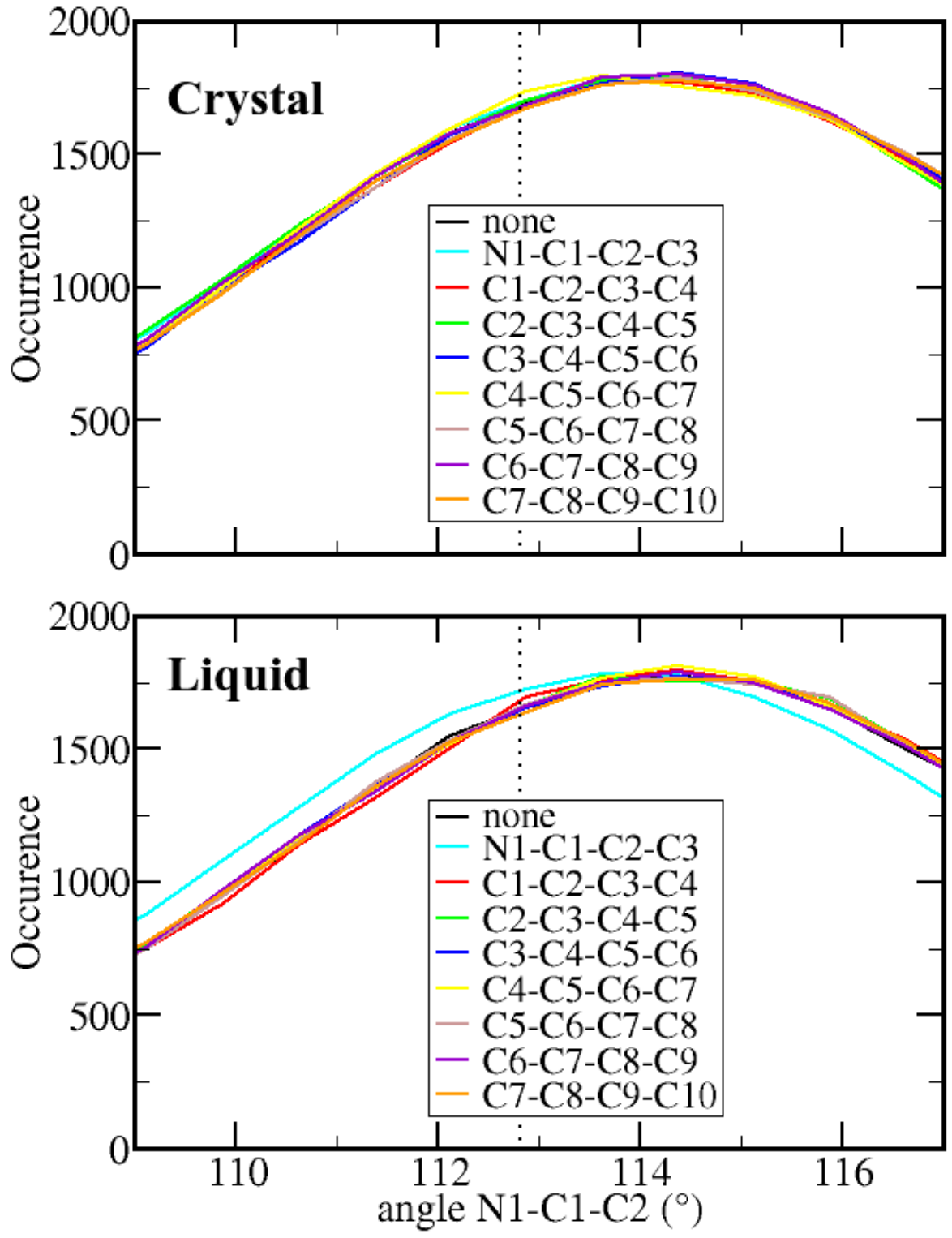

**Figure S21** Distribution of the intramolecular N1-C1-C2 angle for the crystal (top) and for the liquid (bottom) without restraint (black curve) and with restraints at different positions (colored curves). The vertical dotted line indicates the position of the minimum of the harmonic potential employed for the angle.

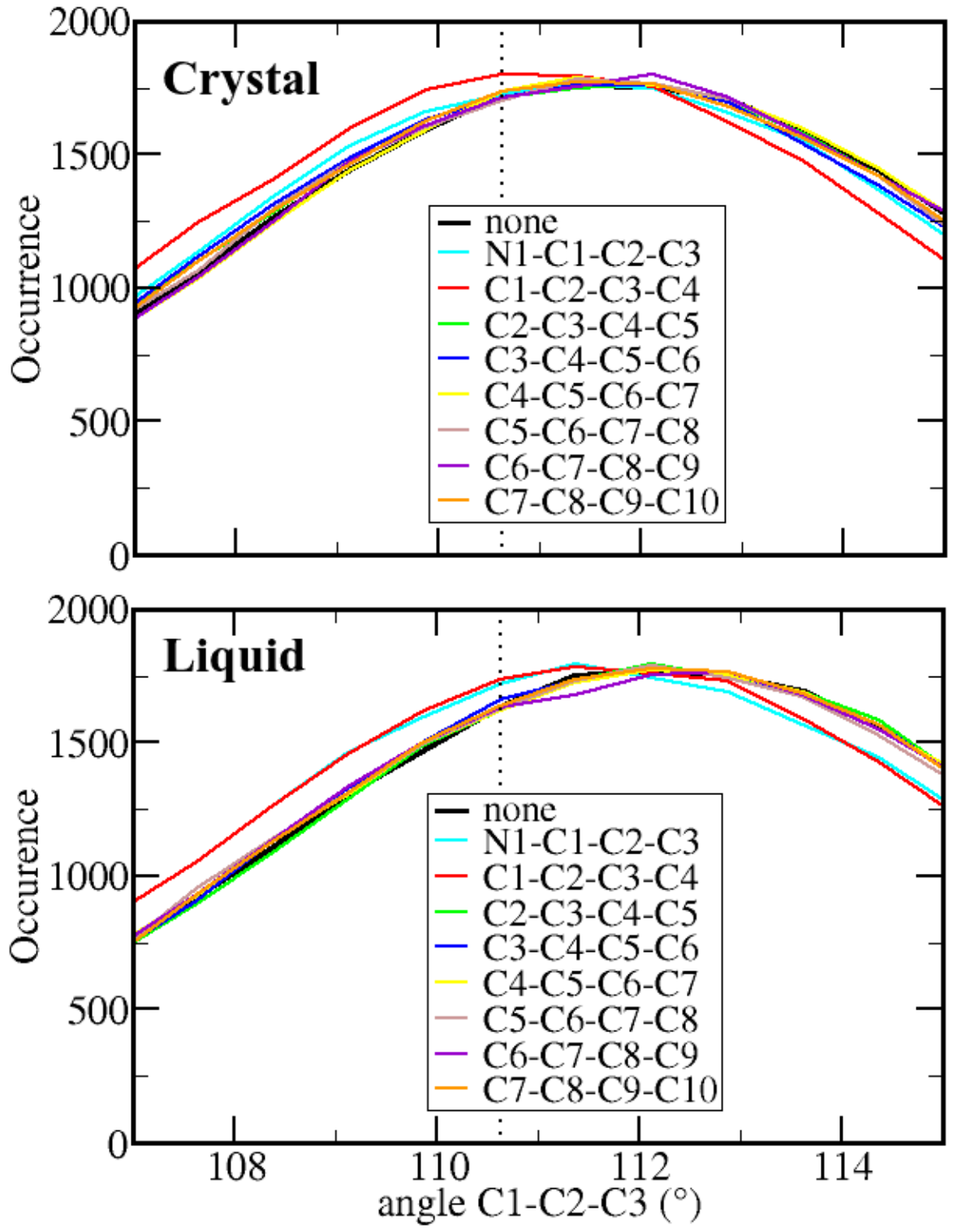

**Figure S22** Distribution of the intramolecular C1-C2-C3 angle for the crystal (top) and for the liquid (bottom) without restraint (black curve) and with restraints at different positions (colored curves). The vertical dotted line indicates the position of the minimum of the harmonic potential employed for the angle.

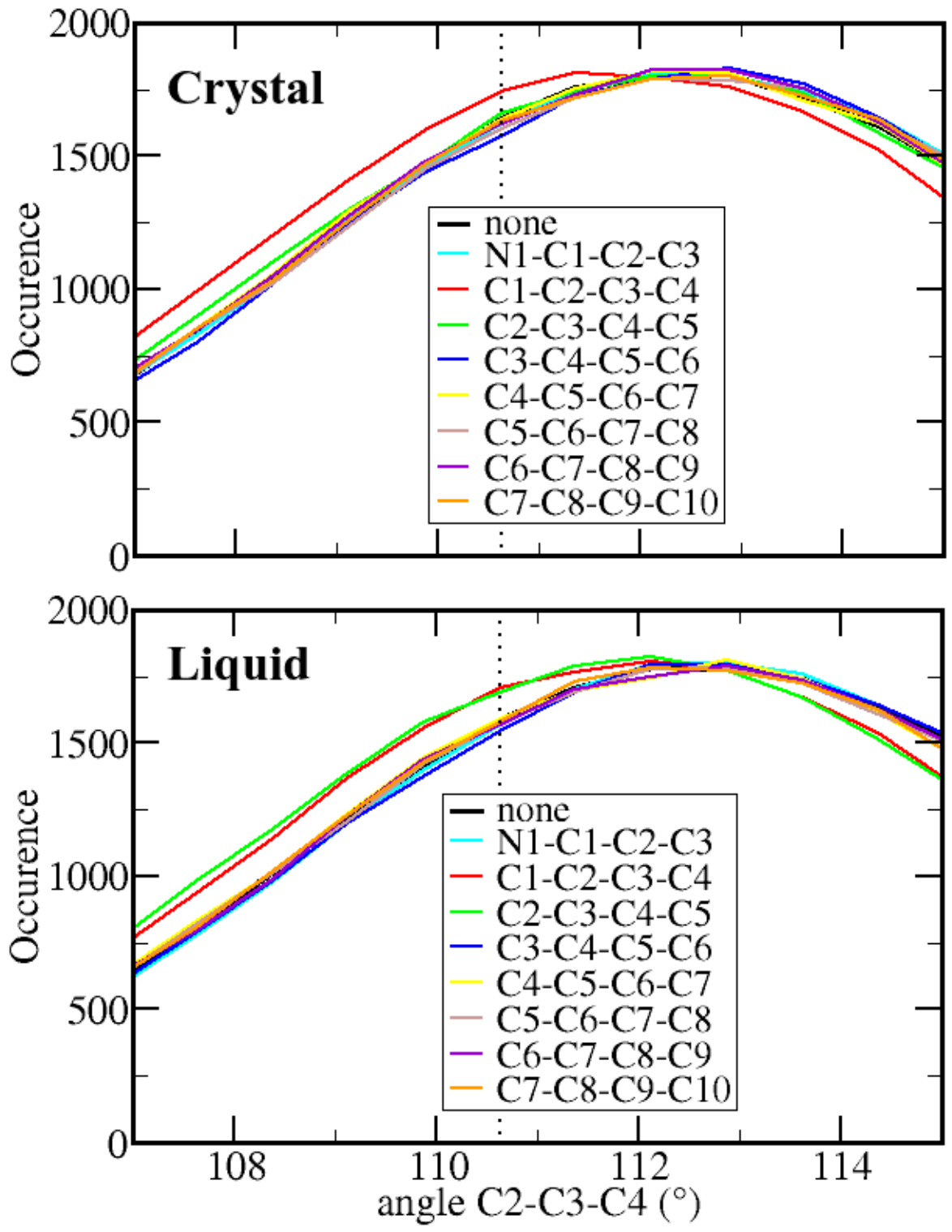

**Figure S23** Distribution of the intramolecular C2-C3-C4 angle for the crystal (top) and for the liquid (bottom) without restraint (black curve) and with restraints at different positions (colored curves). The vertical dotted line indicates the position of the minimum of the harmonic potential employed for the angle.

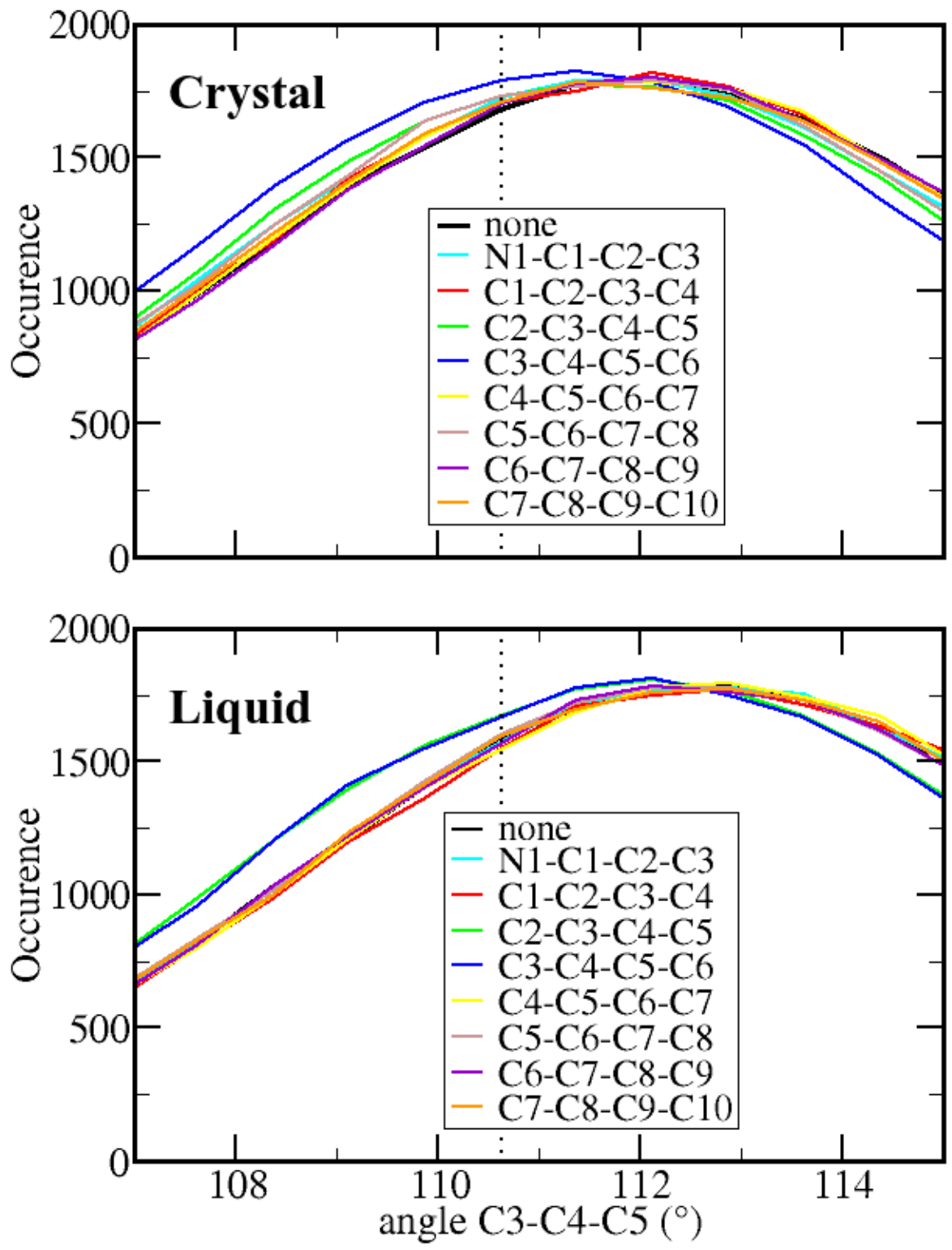

**Figure S24** Distribution of the intramolecular C3-C4-C5 angle for the crystal (top) and for the liquid (bottom) without restraint (black curve) and with restraints at different positions (colored curves). The vertical dotted line indicates the position of the minimum of the harmonic potential employed for the angle.

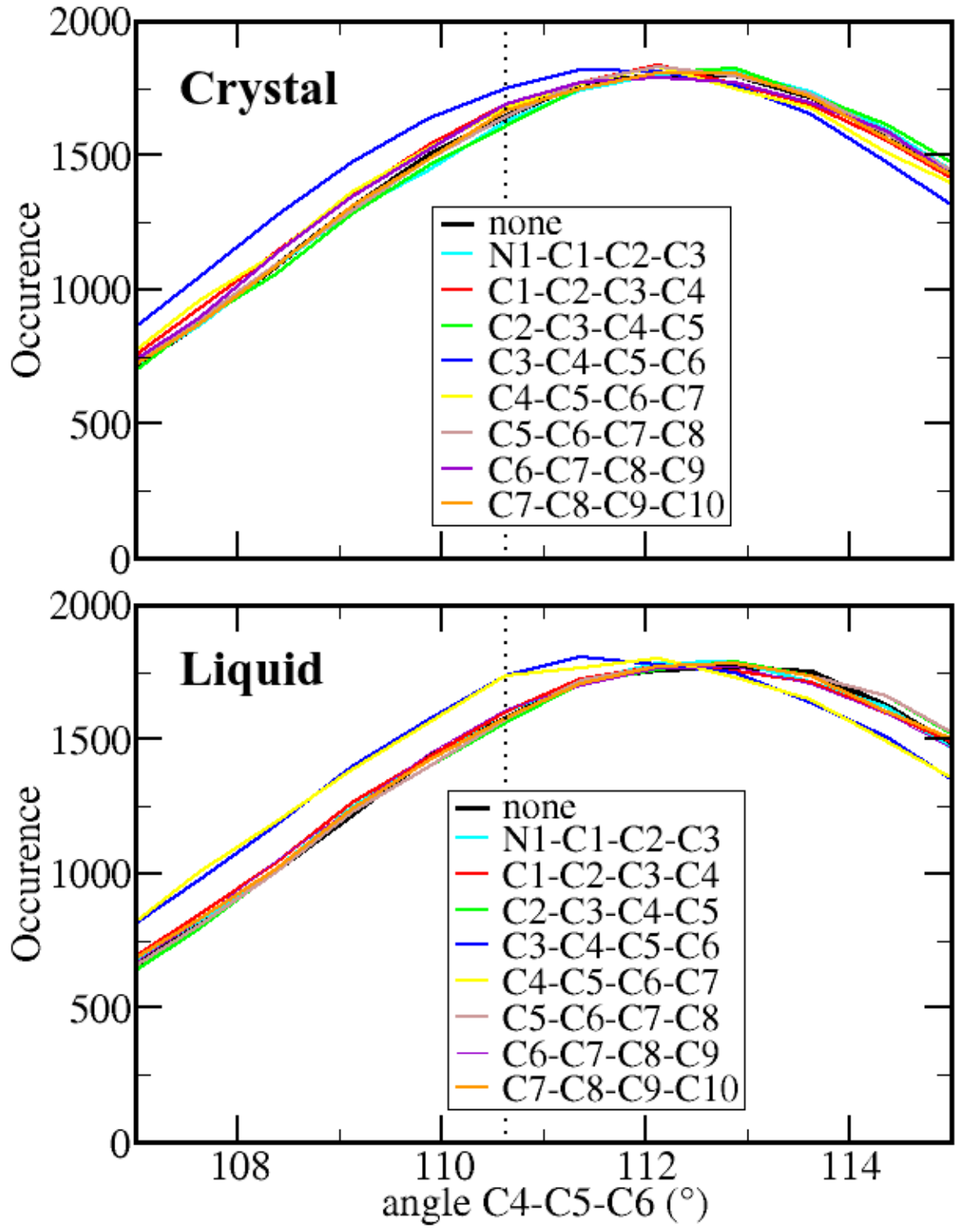

**Figure S25** Distribution of the intramolecular C4-C5-C6 angle for the crystal (top) and for the liquid (bottom) without restraint (black curve) and with restraints at different positions (colored curves). The vertical dotted line indicates the position of the minimum of the harmonic potential employed for the angle.

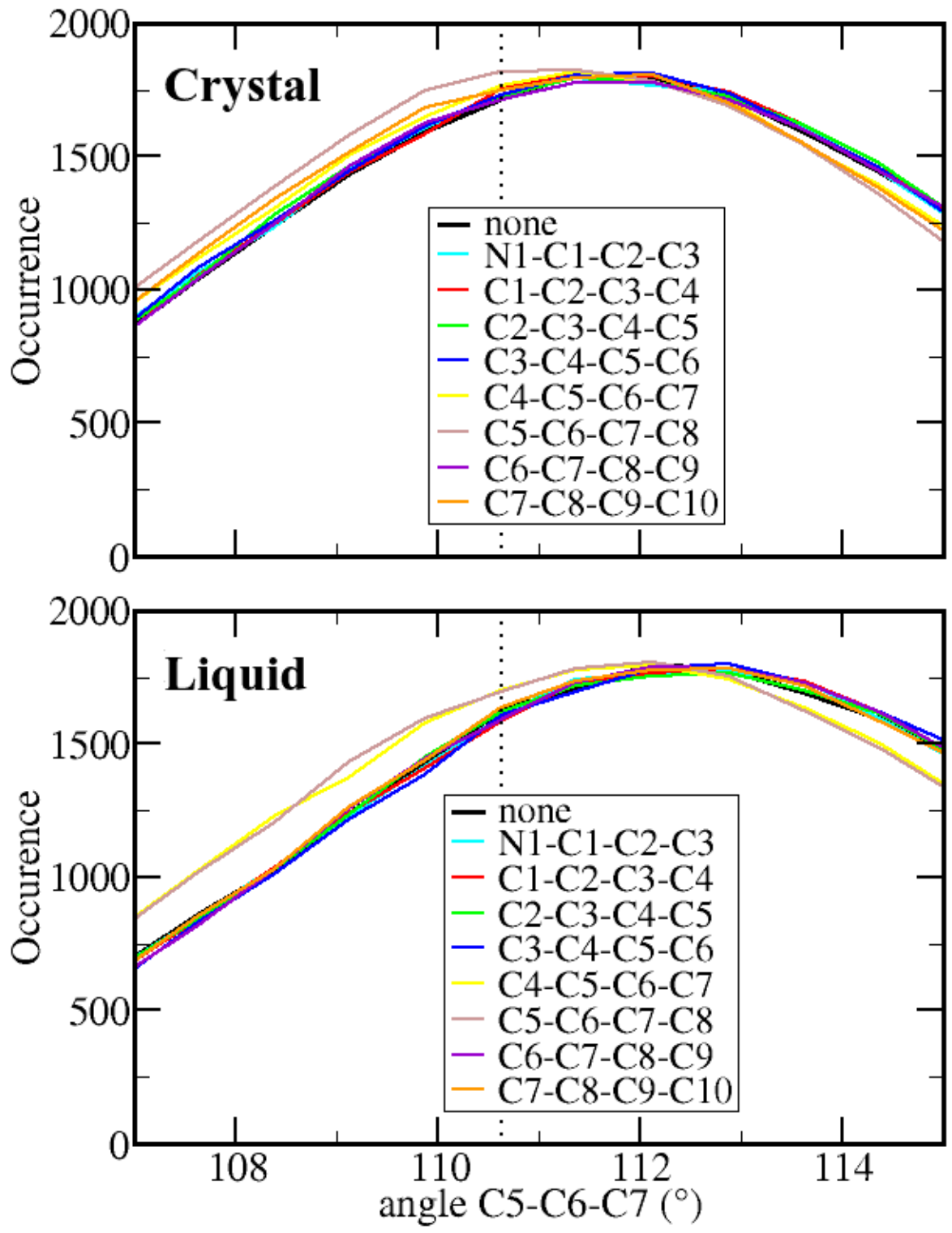

**Figure S26** Distribution of the intramolecular C5-C6-C7 angle for the crystal (top) and for the liquid (bottom) without restraint (black curve) and with restraints at different positions (colored curves). The vertical dotted line indicates the position of the minimum of the harmonic potential employed for the angle.

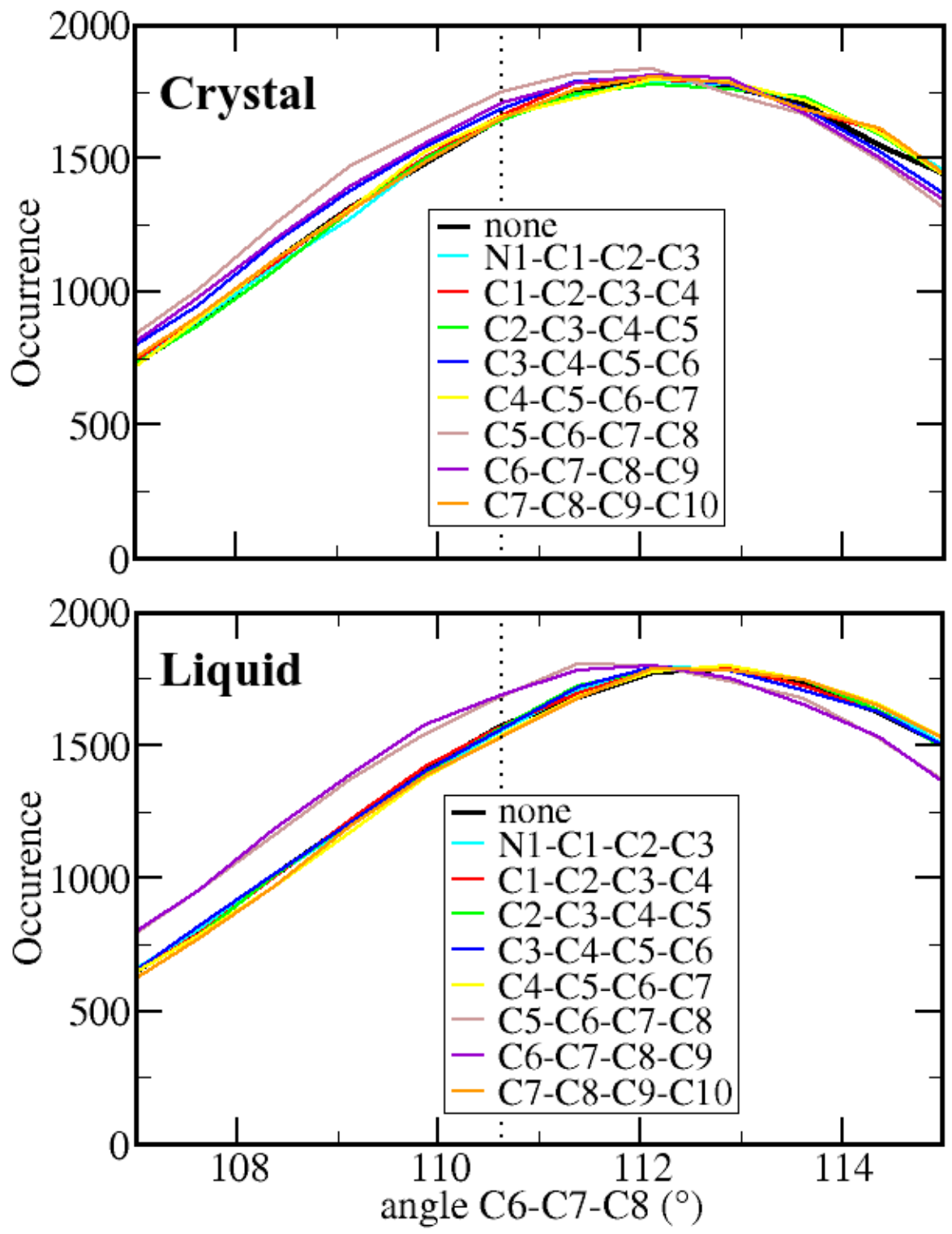

**Figure S27** Distribution of the intramolecular C6-C7-C8 angle for the crystal (top) and for the liquid (bottom) without restraint (black curve) and with restraints at different positions (colored curves). The vertical dotted line indicates the position of the minimum of the harmonic potential employed for the angle.

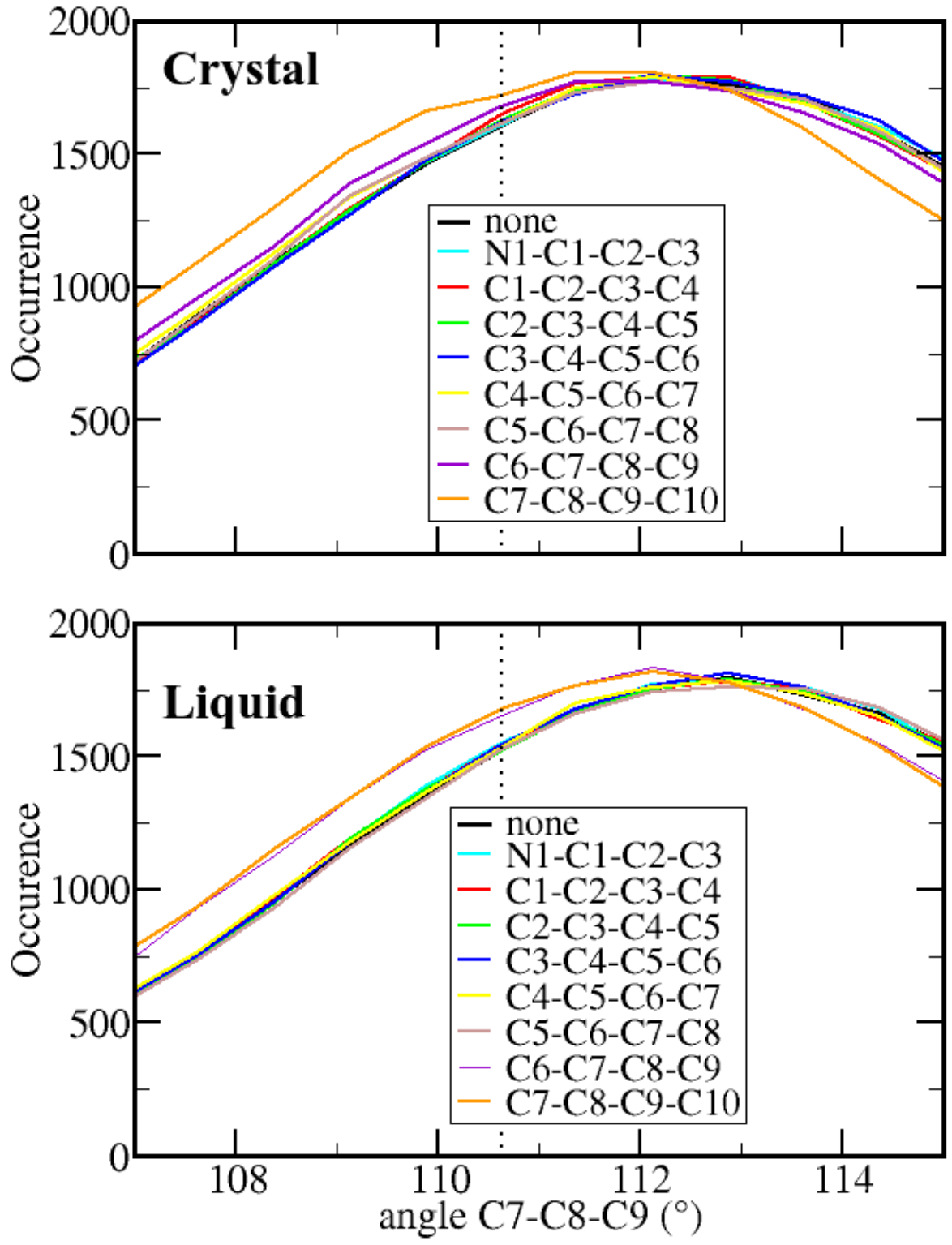

**Figure S28** Distribution of the intramolecular C7-C8-C9 angle for the crystal (top) and for the liquid (bottom) without restraint (black curve) and with restraints at different positions (colored curves). The vertical dotted line indicates the position of the minimum of the harmonic potential employed for the angle.

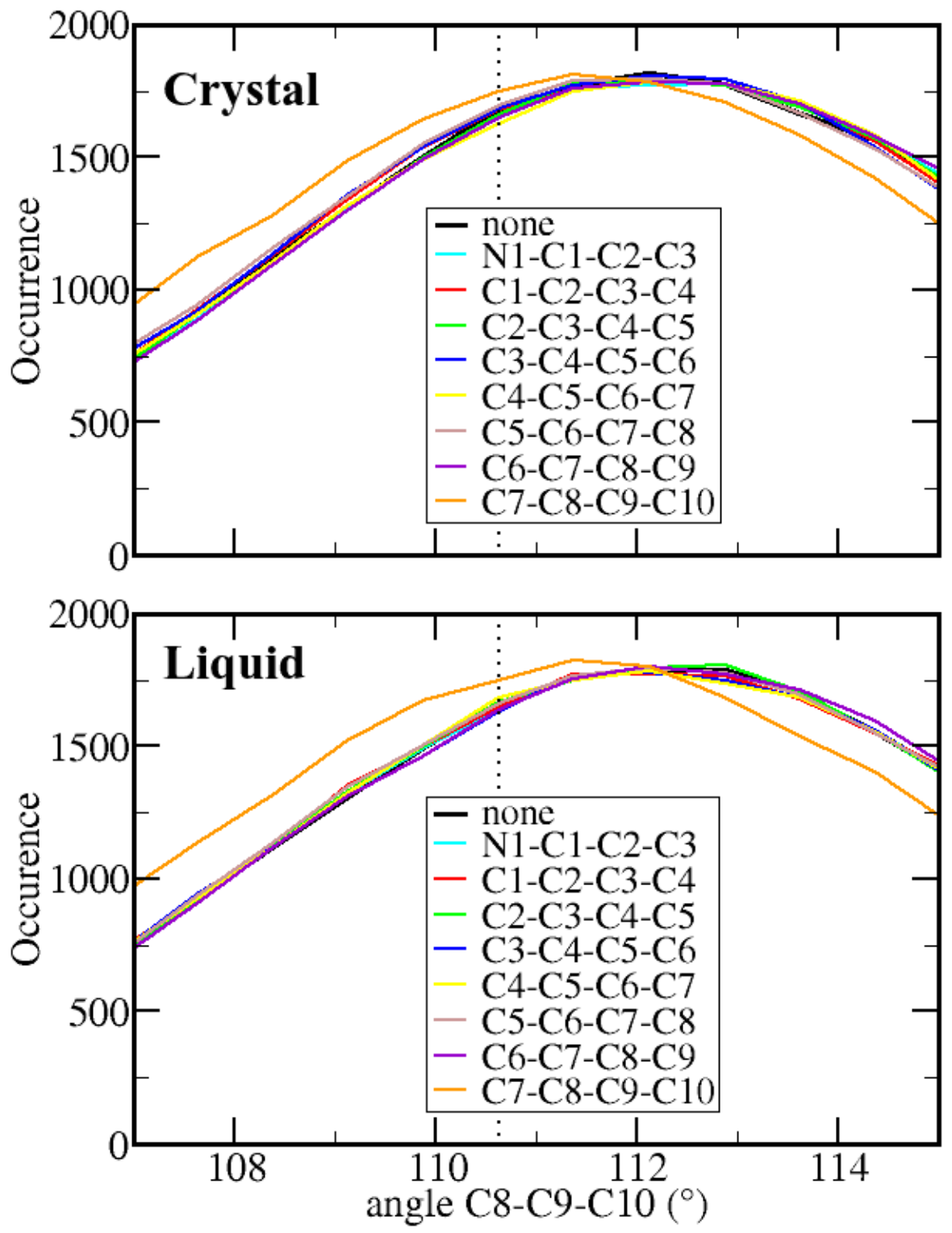

**Figure S29** Distribution of the intramolecular C8-C9-C10 angle for the crystal (top) and for the liquid (bottom) without restraint (black curve) and with restraints at different positions (colored curves). The vertical dotted line indicates the position of the minimum of the harmonic potential employed for the angle.